\documentclass[sigconf]{acmart} 

\AtBeginDocument{%
  \providecommand\BibTeX{{%
    \normalfont B\kern-0.5em{\scshape i\kern-0.25em b}\kern-0.8em\TeX}}}

\copyrightyear{2022} 
\acmYear{2022} 
\setcopyright{acmcopyright}
\acmConference[CIKM '22]{Proceedings of the 31st ACM International Conference on Information and Knowledge Management}{October 17--21, 2022}{Atlanta, GA, USA} 
\acmBooktitle{Proceedings of the 31st ACM International Conference on Information and Knowledge Management (CIKM '22), October 17--21, 2022, Atlanta, GA, USA} 
\acmPrice{15.00} 
\acmDOI{10.1145/3511808.3557343} 
\acmISBN{978-1-4503-9236-5/22/10}

\usepackage{bm}
\usepackage{caption}
\usepackage{graphicx}
\usepackage{subfigure}
\usepackage{amsmath}
\usepackage{booktabs}
\usepackage{threeparttable}
\usepackage{multirow}
\usepackage{enumitem}
\usepackage{xcolor}
\usepackage{mathtools}
\usepackage[ruled]{algorithm2e}

\begin{document}

\title{Hard Negatives or False Negatives: Correcting Pooling Bias in Training Neural Ranking Models}

\author{Yinqiong Cai}
\affiliation{
 \institution{CAS Key Lab of Network Data Science and Technology, ICT, CAS}
 \institution{University of Chinese Academy of Sciences}
 \city{Beijing}
 \country{China}
}
\email{caiyinqiong18s@ict.ac.cn}

\author{Jiafeng Guo}
\authornote{Jiafeng Guo is the corresponding author.}
\affiliation{
 \institution{CAS Key Lab of Network Data Science and Technology, ICT, CAS}
 \institution{University of Chinese Academy of Sciences}
 \city{Beijing}
 \country{China}
}
\email{guojiafeng@ict.ac.cn}

\author{Yixing Fan}
\affiliation{
 \institution{CAS Key Lab of Network Data Science and Technology, ICT, CAS}
 \institution{University of Chinese Academy of Sciences}
 \city{Beijing}
 \country{China}
}
\email{fanyixing@ict.ac.cn}

\author{Qingyao Ai}
\affiliation{
 \institution{Dept. CS\&T, Beijing National Research Center for Information Science and
Technology, Tsinghua University}
 \city{Beijing}
 \country{China}
}
\email{aiqy@tsinghua.edu.cn}

\author{Ruqing Zhang}
\affiliation{
 \institution{CAS Key Lab of Network Data Science and Technology, ICT, CAS}
 \institution{University of Chinese Academy of Sciences}
 \city{Beijing}
 \country{China}
}
\email{zhangruqing@ict.ac.cn}

\author{Xueqi Cheng}
\affiliation{
 \institution{CAS Key Lab of Network Data Science and Technology, ICT, CAS}
 \institution{University of Chinese Academy of Sciences}
 \city{Beijing}
 \country{China}
}
\email{cxq@ict.ac.cn}
\renewcommand{\shortauthors}{Yinqiong Cai et al.}

\begin{abstract}
Neural ranking models (NRMs) have become one of the most important techniques in information retrieval (IR). 
Due to the limitation of relevance labels, the training of NRMs heavily relies on negative sampling over unlabeled data. 
In general machine learning scenarios, it has shown that training with hard negatives (i.e., samples that are close to positives) could lead to better performance. 
Surprisingly, we find opposite results from our empirical studies in IR. 
When sampling top-ranked results (excluding the labeled positives) as negatives from a stronger retriever, the performance of the learned NRM becomes even worse. 
Based on our investigation, the superficial reason is that there are more false negatives (i.e., unlabeled positives) in the top-ranked results with a stronger retriever, which may hurt the training process; The root is the existence of \textit{pooling bias} in the dataset constructing process, where annotators only judge and label very few samples selected by some basic retrievers.
Therefore, in principle, we can formulate the false negative issue in training NRMs as learning from labeled datasets with pooling bias. 
To solve this problem, we propose a novel Coupled Estimation Technique (CET) that learns both a relevance model and a selection model simultaneously to correct the pooling bias for training NRMs.
Empirical results on three retrieval benchmarks show that NRMs trained with our technique can achieve significant gains on ranking effectiveness against other baseline strategies.
\end{abstract}



\maketitle

\section{Introduction}

To balance the effectiveness and efficiency, modern information retrieval (IR) systems generally employ a multi-stage architecture where a \textit{retriever} (e.g., BM25~\cite{robertson2009probabilistic}, ANCE~\cite{xiong2020approximate}) is firstly used to quickly retrieve a few potentially relevant documents from a large collection and then a \textit{ranker} (e.g., monoBERT~\cite{nogueira2019passage}, CEDR~\cite{macavaney2019cedr}) is employed to further analyze and re-rank these documents for precise ranking~\cite{matveeva2006high}.  
Recently, deep learning techniques have been widely applied to construct the rankers, and these neural ranking models (NRMs) have shown advanced performance in modern IR systems~\cite{guo2020deep}.  
Similar to other deep learning models, most existing NRMs are data-hungry.
Thus, with limited relevance data in IR, i.e., sparse-labeled documents and a large number of unlabeled documents for each query, the training of NRMs heavily relies on negative sampling over unlabeled data~\cite{nguyen2016ms, craswell2020overview}.
To facilitate NRMs training, different negative sampling strategies have been explored~\cite{huang2013learning, macavaney2019cedr}.
Among them, the most popular strategy is to sample negatives from the top-ranked results (excluding the labeled positives) returned by the retriever~\cite{macavaney2019cedr,ma2021prop}.

\begin{figure}[!t]
\setlength{\abovecaptionskip}{5pt}
\setlength{\belowcaptionskip}{-0.5cm}
\centering
\includegraphics[scale=0.38]{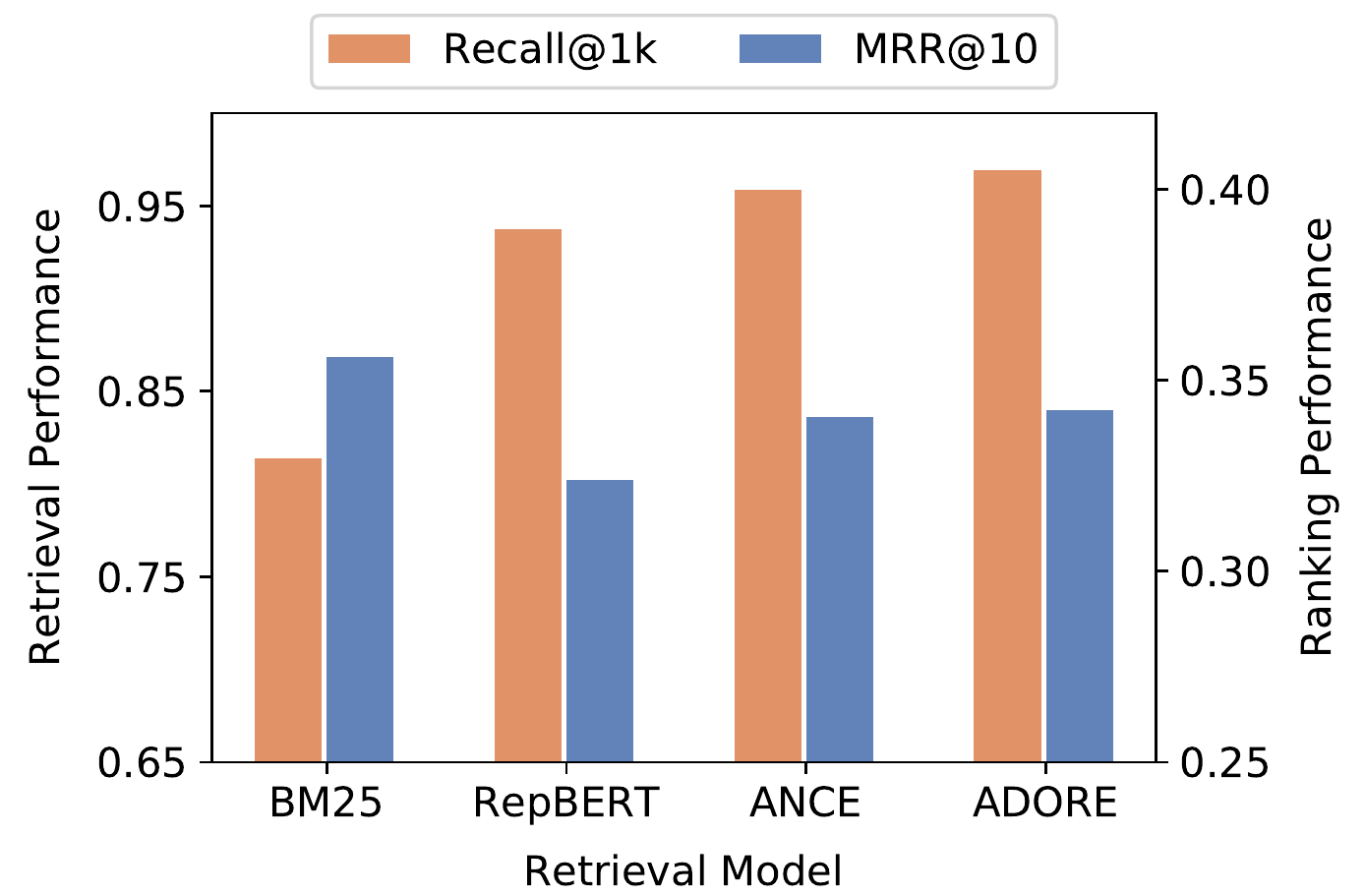}
\caption{Performance on the MS MARCO Passage Ranking task for four different retrievers (Recall@1k) and BERT-base rankers~\cite{nogueira2019passage} (MRR@10). Negatives for rankers training are randomly sampled from top-1000 passages returned by corresponding retrievers.} 
\label{introduction}                                  
\end{figure}

Compared to random negative samples, studies in the machine learning community have shown that training with hard negative samples (i.e., negatives that are similar to positives) can better improve the performance of deep learning models~\cite{robinson2020contrastive}.
In IR scenarios, such hard negatives can be easily obtained from the top-ranked results of a stronger retriever, such as RepBERT~\cite{zhan2020repbert}, ANCE~\cite{xiong2020approximate}, and ADORE~\cite{zhan2021optimizing}, which has more sophisticated structure and much better recall capacity than traditional statistical retrievers such as BM25.
Surprisingly, as shown in our empirical experiments on the MS MARCO Passage Ranking task (see Figure~\ref{introduction}), directly sampling hard negatives from an improved retriever for NRMs training could lead to inferior ranking performance. 
For example, RepBERT~\cite{zhan2020repbert} is a better retriever than BM25 (e.g., the Recall@1k of RepBERT is 15\% better than BM25), but the BERT-base ranker~\cite{nogueira2019passage} trained with negatives sampled from RepBERT is 9\% worse than the same ranker trained with samples from BM25 on MRR@10.

To investigate the cause of this phenomenon, we trace back to the hard negative samples from different retrievers for NRMs training.
Interestingly, we observe that the training data constructed with stronger retrievers is more likely to be riddled with false negatives (i.e., unlabeled positives).
This observation is consistent with the previous study~\cite{arabzadeh2021shallow} which found that the top-ranked passages returned by the retrievers on the MS MARCO leaderboard often appear as good as, or even superior to, the qrels (i.e., positive samples labeled by annotators explicitly).
Without proper treatment, the increasing rate of false negatives in the training data would inevitably hurt the NRMs training process~\cite{ding2020rocketqa, zhang2021dml}.
To this end, one may image that a direct solution to this problem is to apply some heuristic or statistical methods~\cite{ding2020rocketqa} to classify the false negatives from the negative samples during training.

However, if we take a deeper look at the construction of the training data, we may find that the above false negative problem is actually related to the \textit{pooling bias}\footnote{Note that the pooling bias in this study focuses on the effect on model training, not the evaluation.} in the labeling process in IR. 
The pooling bias happens when annotators are instructed to judge and label documents within a small set pooled by some basic retrievers (e.g., Bing for the MS MARCO dataset construction~\cite{nguyen2016ms}). 
In other words, the labeling process is \textit{biased} by basic retrievers one have chosen. 
Under this setting, all the documents outside the pooled set are left unjudged, leading to the potential existence of unlabeled positives. 
When these unlabeled positives are sampled as negatives, the above false negative problem emerges. 
The stronger the sampler is, the severer the problem becomes. 
Therefore, we argue that the root of the above false negative problem is the pooling bias in IR, and we refer more detailed discussions to Section~\ref{sec:pooling_bias}.

In this way, we no longer treat the false negative issue in training NRMs as a simple classification problem, but rather formulate it as a learning problem from the biased dataset.
Inspired by previous studies on unbiased learning-to-rank~\cite{wang2016learning, ai2018unbiased, joachims2017unbiased}, we propose a Coupled Estimation Technique (CET) to solve the NRM training problem from the labeled dataset with pooling bias. 
Specifically, CET attempts to train a selection model to estimate the selection propensity of each document, and a relevance model to estimate the true relevance degree of each document given the query. 
We demonstrate that these two models can be trained in an unbiased way with the help of each other over the biased dataset. 
Therefore, CET employs a coupled learning procedure to learn both models simultaneously.
Based on this, we are able to weaken false negatives and derive high quality hard negatives for NRMs training.

To evaluate the effectiveness of CET, we conduct extensive experiments on three retrieval benchmarks.  
Empirical results demonstrate that NRMs learned with CET can achieve significantly better performance over that learned with state-of-the-art techniques to address the false negative issue.
Moreover, CET is shown to be effective in training different rankers with hard negatives from a variety of retrievers.
In addition, unlike baseline techniques that suffer from high sensitivity w.r.t hyperparameters (which are usually set according to the prior information about the distribution of false negatives in the dataset), CET is demonstrated to be more stable and robust in general.



\section{Related Work}
In this section, we present topics related to our work: bias in information retrieval and negative sampling strategies in IR.

\subsection{Bias in Information Retrieval}
Data for IR tasks are usually collected in two ways: explicit labeling by annotators (i.e., labeled data) and implicit feedback from users (i.e., click data). 
We review typical bias problems in them. 

\textbf{Bias in labeled data.}
Early IR systems are built on the relevance data constructed with a pooling process~\cite{spark1975report}: existing retrieval models are firstly used to get a document pool, and then annotators are instructed to label their relevance.
The idea behind the pooling technique is to find enough relevant documents such that the relevance data is sufficiently complete and unbiased when only partial documents are judged~\cite{clarke2009overview}.  
However, with the growth of the document set size and the development of IR techniques, the assumption of approximately complete labeling becomes invalid~\cite{clarke2004overview, arabzadeh2021shallow, thakur2021beir}.
In this case, the pooling bias problem is identified by researchers, which would underestimate the effectiveness of new IR systems with standard evaluation~\cite{buckley2006bias}.
For example, \citet{webber2009score} proposed to estimate the degree of bias against a new retriever to adjust the evaluation score.
Differing from the above early works, recently researchers realize the detrimental effect of pooling bias to IR models training~\cite{thakur2021beir, arabzadeh2021shallow}, where unlabeled positives, called false negatives in this work, would make the learned models biased and hurt their performance.
However, these pioneers only discussed this problem preliminarily and intuitively.
In this work, we will show a exhaustive analysis and give a formal definition to it.

\textbf{Bias in click data.}
Due to the cost of annotators labeling, click data is easily collected and has developed to be a critical resource for IR models training. 
However, various bias problems, such as position bias~\cite{joachims2017unbiased}, selection bias~\cite{wang2016learning, ovaisi2020correcting} and presentation bias~\cite{yue2010beyond}, make it difficult to be directly leveraged as training data. 
For example, position bias is caused by the position where a document is displayed to users, making higher ranked documents more likely to be clicked.
~\citet{ovaisi2020correcting} claimed that users rarely have the chance or energy to check about all documents in the lists.
In this case, lower-ranked relevant documents have zero probability of being clicked, thus leading to the selection bias.
To make full use of click data, studies on unbiased learning-to-rank~\cite{ai2018ULTR} have attracted a lot of attention.
Among them, the counterfactual estimation technique from causal inference, such as inverse propensity weighting (IPW)~\cite{wang2016learning, joachims2017unbiased, ai2018unbiased} and Heckman correction~\cite{ovaisi2020correcting}, is used to solve these bias problems in click data.
Different from these studies, in this work, we focus on the bias in labeled data, where we do not possess any prior knowledge on the labeling process and cannot explicitly quantify the bias distribution with user studies.


\subsection{Negative Sampling Strategies in IR} \label{negative_sampling}
In practice, training data for IR models usually consist of sparse-labeled documents  for a set of queries and a large number of unlabeled documents.
Among the labeled documents, there are often limited~\cite{clarke2009overview,qin2013introducing} or even no~\cite{nguyen2016ms, qiu2022dureader_retrieval} samples with explicit negative labels. 
Therefore, negative sampling over unlabeled documents is usually necessary for training IR models. 

To find informative negatives for models training, many methods have been explored in IR.
Random sampling from the document set is the simplest and most direct way to get negatives, which is a widely used strategy in existing works~\cite{karpukhin2020dense, huang2013learning}. 
Nonetheless, such approach is sub-optimal because random negatives have been proven to be too easy to learn effective models for generalizing to sophisticated testing cases~\cite{lakshman2021embracing, robinson2020contrastive}.   
Instead, negative samples that are more difficult to be distinguished from positive ones are more desired. 
There have been studies showing that these \textit{hard} negative samples could improve model generalization and accelerate convergence~\cite{oh2016deep, schroff2015facenet}. 
However, efficiently identifying such informative negative samples emerges as a challenge since it is computationally infeasible to examine all possible samples. 
Among current works in the IR field, the most commonly used hard negatives are the top-ranked documents of a strong retriever~\cite{xiong2020approximate, ren2021rocketqav2, zhang2021adversarial}.

Recently, with the prevalence of pre-training methods, the availability of stronger retrievers~\cite{guo2022semantic} shows new potentials to provide hard negatives for IR models training.
Meanwhile, researchers begin to realize the severity of false negatives for neural retrievers and rankers training~\cite{ding2020rocketqa, gao2021rethink, arabzadeh2022shallow, mackenzie2021sensitivity}. 
For example, ~\citet{ding2020rocketqa} found that 70\% of the top-ranked passages returned by their retriever are actually positives or highly relevant in MS MARCO~\cite{nguyen2016ms}.
To address this problem, there have been several studies trying to filter false negatives during the negative sampling process with some heuristic methods.
For example, RocketQA~\cite{ding2020rocketqa} proposed to train another ranking model in advance to evaluate relevance scores for all unlabeled documents and sets a hard threshold empirically to remove those potentially false negatives. 
Additionally, RANCE~\cite{prakash2021learning} presented a special sampling technique to filter false negatives. It estimates the negative sampling distribution relying on a separate dataset with complete relevance labeling, which, unfortunately, is usually not available in practice.
These two works are highly related to this study, and we implement two baselines based on them respectively to compare with our method.

\section{Problem Description}
In this section, we briefly review the general training paradigm of NRMs on labeled datasets and introduce the pooling bias in detail. We show how this pooling bias leads to the false negative problem in negative sampling as observed in the Introduction Section. Based on these analysis, we formulate the false negative issue in training NRMs as a learning problem over biased labeled datasets.

\subsection{Training NRMs on Labeled Dataset}   \label{training_on_labeled_data}
We begin by introducing the general setting of NRMs learning with fully labeled data, also referred to as the \textbf{Full-Information Setting}~\cite{joachims2017unbiased}. Ideally, given a query $q \in \mathcal{Q}$, the relevance $r_i$ of each document $d_i \in \mathcal{D}$ is known beforehand. 
For simplicity, we assume that the relevance is binary (i.e., $r \in \{0, 1\}$) and one can easily extend it to the multi-level relevance case.
In practice, the pairwise learning setting tends to be more effective for ranking models training and has been widely adopted~\cite{burges2010ranknet, Unbiased2019Hu}.
Thus, here we consider that the ranking model $R$ is defined on a query-document pair, and the loss function is defined on a triple $(q, d_i, d_j)$.
Let $r_{i}^{+}$ and $r_{j}^{-}$ represent that $d_i$ is relevant (i.e., $r_i=1$) and $d_j$ is irrelevant (i.e., $r_j=0$) for $q$. Let $x_i$ and $x_j$ denote feature vectors from $d_i$ and $d_j$ as well as $q$.
The risk function for NRMs learning is defined as:
\begin{equation}
\small
\label{full_relevance_loss}
\mathcal{R}_{\mathit{full}}(R) = \int L\left(R(x_i), R(x_j)\right) \ dP(x_i, r_i^+, x_j, r_j^-).
\end{equation}
Given the fully labeled dataset, the ranker $R$ can be learned by minimizing the empirical risk function as follows:
\begin{equation}
\small
\label{full_relevance_model}
\hat{R}_{\mathit{full}} = \arg \min_{R}\sum_{\mathit{q \in \mathcal{Q}}} \sum_{\mathit{(d_{i}, d_{j}) \in \mathcal{D}}}\!L\left(R(x_{i}), R(x_{j})\right),
\end{equation}
where the document pair $(d_i, d_j)$ denotes that $d_i$ is a document with $r_i=1$ and $d_j$ is a document with $r_j=0$ for the query $q$.

However, in practice, it is impossible to judge and label all the documents in the corpus for each query.
To reduce the annotation effort, existing methods usually apply some basic retrievers to select a small set of documents for labeling~\cite{spark1975report}.
That is, for each query, only a few selected documents can be judged and labeled by annotators.
In this case, NRMs are trained under a \textbf{Partial Information Setting}~\cite{ovaisi2020correcting}, where only a part of relevance information for each query is available and most remains unobserved.
With this labeled dataset for training NRMs, except for the very few labeled documents, all unlabeled documents are usually deemed to be negatives (i.e., irrelevant). Then the risk function and minimization of empirical risk function could be reformalized as:
\begin{equation}
\small
\label{partial_relevance_loss}
\begin{aligned}
\mathcal{R}_{\mathit{part}}(R) = & \int L\left(R(x_i), R(x_j)\right) \ dP(x_i, l_i^+, x_j, l_j^-) \\
&+ \int L\left(R(x_i), R(x_j)\right) \ dP(x_i, l_i^+, x_j, ul_j),
\end{aligned}
\end{equation}
\begin{equation}
\small
\label{partial_relevance_model}
\hat{R}_{\mathit{part}} =\arg \min_{R} \sum_{\mathit{q \in \mathcal{Q}}} \sum_{\mathit{\substack{d_{i}\in \mathcal{D}^{+} \\ d_{j} \in \mathcal{D}^{-} \cup \mathcal{D}^{\mathit{ul}}}}} \! L\left(R(x_i), R(x_j)\right),
\end{equation}
where $l_{i}^{+}$ and $l_{j}^{-}$ denote that $d_i$ is labeled as relevant (i.e., $l_i=1$) and $d_j$ is labeled as irrelevant (i.e., $l_j=0$) for $q$ respectively, $ul_j$ denotes that $d_j$ is an unlabeled document for $q$, $\mathcal{D} = \mathcal{D}^{+} \cup \mathcal{D}^{-} \cup \mathcal{D}^{\mathit{ul}}$, and these three document sets are disjointing.
In practice, the labeled negatives (i.e., $\mathcal{D}^{-}$) are often missing~\cite{nguyen2016ms, qiu2022dureader_retrieval}, thus it is essential to utilize documents in $\mathcal{D}^{\mathit{ul}}$ for learning effective NRMs~\cite{huang2013learning, macavaney2019cedr,ma2021prop}.

\subsection{Pooling Bias in Labeled Data} \label{sec:pooling_bias}

As mentioned in Section~\ref{training_on_labeled_data}, the construction of labeled datasets relies on some basic retrievers, called pooling systems, to select a small set of most promising documents for labeling. The pooling technique is very important for IR since it can significantly reduce manual effort for labeling. Therefore, it has been widely adopted in benchmark construction~\cite{clarke2009overview, nguyen2016ms}.  
While the pooling technique does save labeling efforts, it introduces the undesired bias into labeled data, namely pooling bias here. That is, the labeled data is biased by the preference of basic retrievers used in the pooling process. A direct consequence of this pooling bias is the potential existence of unlabeled positives, i.e., some relevant documents might not be preferred by basic retrievers and thus not be selected for labeling. 

When the unlabeled data are sampled as negatives in training NRMs, those unlabeled positives become false negatives if they are unfortunately selected by the sampler. In general, unlabeled positives are the minority in the massive unlabeled data. Therefore, when the random sampling strategy is adopted, the false negative issue would not be severe. This explains why the traditional training paradigm with negative sampling works for NRMs. However, previous studies have shown that uniformly sampling negatives from $\mathcal{D}^{\mathit{ul}}$ often fails to learn an effective ranking model, since random negatives are too easy to produce effective parameter updates~\cite{lakshman2021embracing, robinson2020contrastive}. Therefore, more advanced studies employ the hard negative sampling strategy, where the top-ranked results (excluding labeled positives) returned by a strong retriever are used as negatives for NRMs training~\cite{huang2013learning,macavaney2019cedr}.
However, such sampling strategies would increase the selection probability of false negatives, since unlabeled positives are more likely to be ranked at top positions by a stronger retriever~\cite{ding2020simplify}. This in turn will hurt NRMs training, which is exactly the observation we have mentioned in Figure~\ref{introduction}.

Based on the above analysis, we can see that the pooling bias is the root of the false negative issue in training NRMs with hard negative sampling strategies. Therefore, directly identifying the false negatives from unlabeled data with some classification models may not touch the heart of the problem. In principle, we can formulate the false negative issue in training NRMs as a learning problem from labeled datasets with pooling bias.

\section{Our Approach} \label{method}
In this section, we introduce our approach in detail. 
We first analyze the problem with a counterfactual learning framework (Section~\ref{bias_analysis}).
Then we propose a coupled estimation technique (Section~\ref{debias_learning}) to solve the NRM learning problem with the biased dataset.

\subsection{Bias Correction Analysis}  \label{bias_analysis}
Existing works on bias correction mostly focus on the click data~\cite{ai2018ULTR,joachims2017unbiased}, where the click of a document is affected by its position, popularity, etc. 
Pooling bias discussed in this study is that the label of a document is affected by whether it is selected during the pooling process. 
Based on the common ground, we follow the inverse propensity weighting (IPW) framework in~\cite{joachims2017unbiased, wang2016learning} to solve the pooling bias in labeled datasets.

In this work, we focus on the typical case in popular large-scale retrieval benchmarks~\cite{nguyen2016ms, qiu2022dureader_retrieval, kwiatkowski2019natural}, where only relevant documents in the judgement pool are labeled by annotators and there is no explicitly labeled negatives (i.e., $\mathcal{D}^{-}=\varnothing$).
As a result, it is unavailable which documents have been selected during the pooling process, which makes it challenging for addressing the pooling bias.
With this labeled dataset, we define three notations for each query-document pair $(q, d)$ in it. Besides the relevance $r$ and labeling $l$ defined in Section~\ref{training_on_labeled_data}, we use $s \!\in\! \{0,1\}$ to indicate whether the document $d$ is selected into the judgement pool for the query $q$.
The merely available information for the labeled dataset is $l$, and here, we consider the noise-free labeling where a document with $l=1$ must be relevant and selected.

Considering that annotators have been well instructed, for the pair $(q, d_i)$ denoted as $x_i$, we have the following premise:
\begin{equation}
\small
\label{premise_1}
P(l_i^+ \mid x_i) = P(s_i^+ \mid x_i) \cdot P(r_i^+ \mid x_i),
\end{equation}
where $s_i^+$ denotes that the document $d_i$ is selected (i.e., $s_i=1$) for $q$.
Besides, for an unlabeled document $d_j$, it is more likely to be irrelevant if it has higher probability being selected, that is,
\begin{equation}
\small
\label{premise_2}
P(r_j^- \mid ul_j, x_j) \propto P(s_j^+ \mid x_j).
\end{equation}
Then, we can learn a bias corrected relevance model with an IPW-based risk function and empirical risk function on the labeled dataset with $\mathcal{D}^{-}=\varnothing$:
\begin{equation}
\small
\label{unbiased_relevance_loss}
\begin{aligned}
\mathcal{R}_{\mathit{IPW}}(R) 
&=\int \frac{L(R(x_i), R(x_j))}{\frac{P(s_i^+ \mid x_i)}{P(s_j^+ \mid x_j)}} \ dP(x_i, l_i^+, x_j, ul_j) \\
&\propto\iint \frac{L(R(x_i), R(x_j))}{\frac{P(l_i^+ \mid x_i)}{P(r_i^+ \mid x_i)} \cdot \frac{1}{P(r_j^- \mid ul_j, x_j)}} \ dP(x_i, l_i^+) \ dP( x_j, ul_j) \\ 
&=\iint \frac{L(R(x_i), R(x_j)) \ dP(x_i, l_i^+) \ dP(x_j, ul_j)}{\frac{P(l_i^+ \mid x_i)}{P(r_i^+ \mid x_i)} \cdot \frac{P(ul_j \mid x_j)}{P(r_j^-, ul_j \mid x_j)}} \\ 
&=\iint \frac{L(R(x_i), R(x_j)) \ dP(x_i, l_i^+) \ dP(x_j, ul_j)}{\frac{P(l_i^+ \mid x_i)}{P(r_i^+ \mid x_i)} \cdot \frac{P(ul_j \mid x_j)}{P(r_j^- \mid x_j)}} \\ 
&=\iint L(R(x_i), R(x_j)) \ dP(x_i, r_i^+) \ dP(x_j, r_j^-) \\ 
&=\int L(R(x_i), R(x_j)) \ dP(x_i, r_i^+, x_j, r_j^-) \\ 
&=\mathcal{R}_{\mathit{full}}(R),
\end{aligned}
\end{equation}
\begin{equation}
\small
\label{unbiased_relevance_model}
\hat{R}_{\mathit{IPW}} =\arg \min_{R} \sum_{q \in \mathcal{Q}} \sum_{\substack{d_i\in \mathcal{D}^{+} \\ d_j \in \mathcal{D}^{\mathit{ul}}}} \frac{L(R(x_i), R(x_j))}{\frac{P(s_i^+ \mid x_i)}{P(s_j^+ \mid x_j)}}.
\end{equation}
For Eq.~(\ref{unbiased_relevance_loss}), it is assumed that the relevance and labeling of $d_i$ are independent from $d_j$ (empirical results show that it works well with this assumption, even one may think that it is not strictly correct), the second step uses Eq.~(\ref{premise_1}) \& (\ref{premise_2}), the third step uses the conditional probability formula, and the fourth step uses the premise that a labeled document must be relevant (i.e., $l^+ \!\Rightarrow r^+$, and thus $r^- \!\Rightarrow ul$ with $\mathcal{D}^{-}=\varnothing$). 
With Eq.~(\ref{unbiased_relevance_loss}), it implies that the model $R$ optimized with the IPW empirical risk minimization on labeled datasets can produce the same relevance model trained with the relevance $r$, 
\begin{equation}
\small
\arg \min_{R} \mathcal{R}_{\mathit{IPW}}(R) = \arg \min_{R} \mathcal{R}_{\mathit{full}}(R).
\end{equation}

To optimize the relevance model with Eq.~(\ref{unbiased_relevance_model}), we need to estimate the selection probability $P(s_i^+ \!\mid\! x_i)$ for each document $d_i \in \mathcal{D}$. Ideally, we can learn a selection model $S$ with the risk function and minimization of empirical risk function:
\begin{equation}
\small
\label{full_selection_loss}
\begin{aligned} 
\mathcal{R}_{\mathit{full}}(S) &=\int L(S(x_i), S(x_j)) \ dP(x_i, s_i^+, x_j, s_j^-), \\
\hat{S}_{\mathit{full}} &=\arg \min_{S}\sum_{q \in \mathcal{Q}} \sum_{(d_i, d_j) \in \mathcal{D}}\!L(S(x_i), S(x_j)),
\end{aligned}
\end{equation}
where $s_i^+$ and $s_j^-$ denote that $d_i$ is selected (i.e., $s_i=1$) and $d_j$ is not selected (i.e., $s_j=0$) by pooling systems for the query $q$, and the document pair $(d_i, d_j)$ denotes that $d_i$ is a document with $s_i\!=\!1$ and $d_j$ is a document with $s_j\!=\!0$ for $q$.
However, similar to the relevance information, the selection information is also partial with $\mathcal{D}^{-}=\varnothing$, since only the labeled relevant document's selection information is available and an unlabeled document may be just irrelevant but not unselected.
Nevertheless, for the unlabeled document $d_j$, it is more likely to be unselected if it has higher relevance probability, that is,
\begin{equation}
\small
\label{premise_3}
P(s_j^- \mid ul_j, x_j) \propto P(r_j^+ \mid x_j).
\end{equation}
Then, similar to the relevance model, the bias corrected selection model could be learned with an IPW-based risk function:
\begin{equation}
\small
\label{unbiased_selection_loss}
\mathcal{R}_{\mathit{IPW}}(S) =\int \frac{L(S(x_i), S(x_j))}{\frac{P(r_i^+ \mid x_i)}{P(r_j^+ \mid x_j)}} \ dP(x_i, l_i^+, x_j, ul_j),
\end{equation}
\begin{equation}
\small
\label{unbiased_selection_model}
\hat{S}_{\mathit{IPW}} =\arg \min_{S} \sum_{q \in \mathcal{Q}} \sum_{\substack{d_i\in \mathcal{D}^{+} \\ d_j \in \mathcal{D}^{ul}}} \frac{L(S(x_i), S(x_j))}{\frac{P(r_i^+ \mid x_i)}{P(r_j^+ \mid x_j)}}.
\end{equation}
Based on Eq.~(\ref{premise_1}) \& (\ref{premise_3}) and the fact that a labeled document must be selected (i.e., $l^+ \!\Rightarrow\! s^+$, and thus $s^- \!\Rightarrow\! ul$ with $\mathcal{D}^{-}=\varnothing$), Eq.~(\ref{unbiased_selection_loss}) is easily proven as Eq.~({\ref{unbiased_relevance_loss}}) and the details are omitted here. 

With Eq.~(\ref{unbiased_relevance_model}) \& (\ref{unbiased_selection_model}), we can find that the key of obtaining a debiased relevance model is to estimate $P(s_i^+ \mid x_i)$, while the key of building a debiased selection model is to estimate $P(r_i^+ \mid x_i)$ for each $(q, d_i)$ in the dataset. 
This indicates that the estimation of selection propensities and relevance scores for documents is coupled with each other, where a better selection model can help to train a better relevance model and vice versa. 
Based on this observation, we propose a coupled estimation technique to train both models simultaneously to correct the pooling bias.

\begin{figure}[!t]
\setlength{\abovecaptionskip}{3pt}
\setlength{\belowcaptionskip}{-0.3cm}
\centering
\includegraphics[scale=0.38]{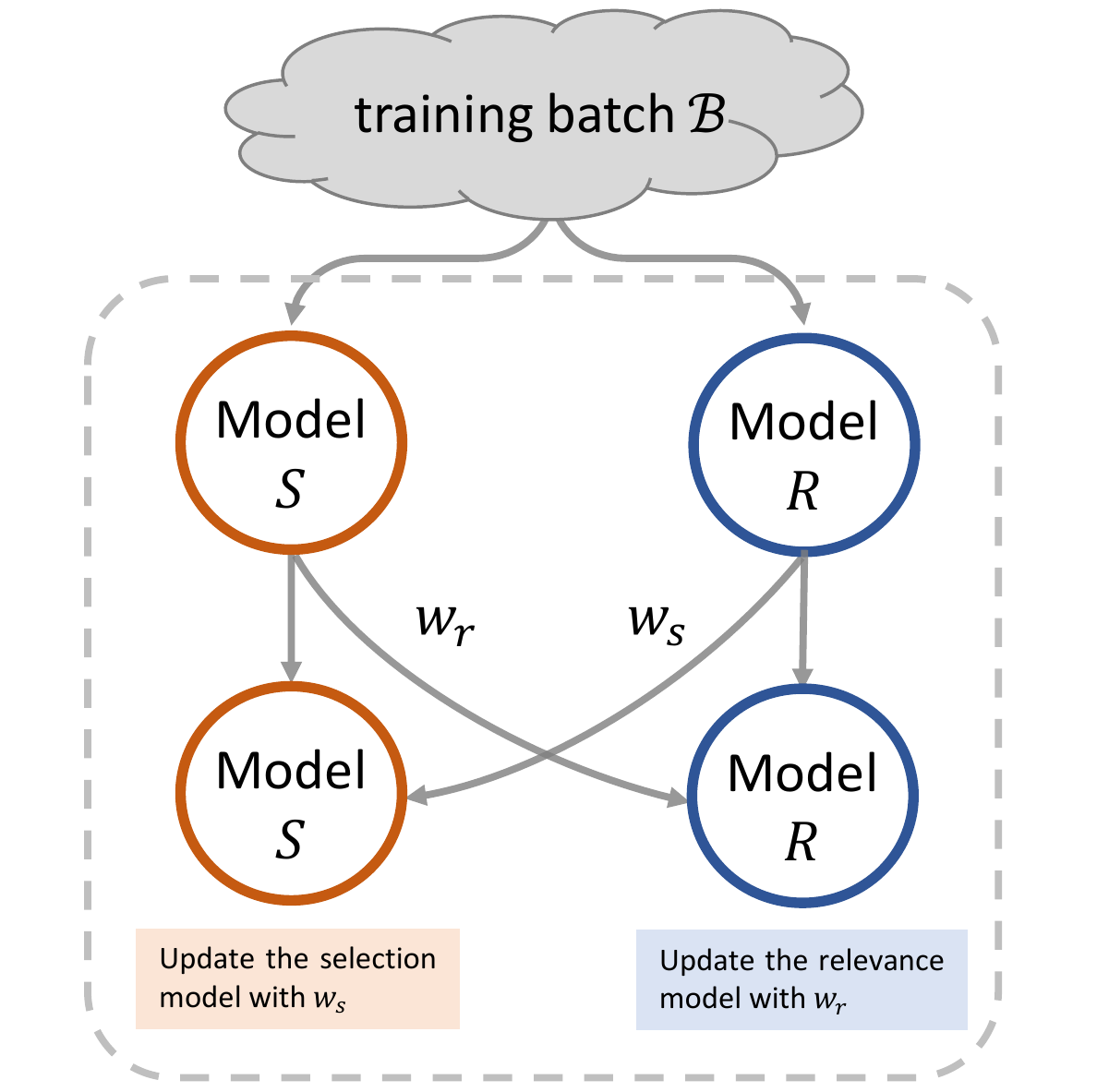}
\caption{The framework of Coupled Estimation Technique (CET). The model $S$ and model $R$ denote the selection model and the relevance model respectively.}
\label{framework}                                  
\end{figure}

\subsection{Coupled Estimation Technique} \label{debias_learning}
Based on the above analysis, we propose a Coupled Estimation Technique (CET) to train NRMs on labeled datasets.
As shown in Fig.~\ref{framework}, the framework of CET includes three components:
\begin{itemize}[leftmargin=*]
\item The relevance model $R$ (implemented with NRMs) that estimates the relevance score for the pair $(q, d)$;
\item The selection model $S$ that estimates the propensity of the document $d$ being selected into the judgement pool for the query $q$ during annotators labeling process;
\item The coupled learning algorithm that optimizes the relevance model and selection model jointly on the labeled dataset.
\end{itemize}

\textbf{Relevance Model.} 
In this work, we implement the relevance model $R$ with a BERT-based architecture parameterized by $\bm{\theta}$.
Following previous studies~\cite{nogueira2019passage}, we concatenate the query $q$ and the document $d$ with special delimiting tokens and feed them into Transformer layers~\cite{vaswani2017attention}. Then, a multi-layer perceptron (MLP) function is applied over the hidden state of the special token [CLS] to obtain the relevance score.
Let $R_{\bm{\theta}}(q, d)$ be the relevance score for the query-document pair $(q, d)$, which can be estimated with,
\begin{equation}
\label{relevance_score}
R_{\mathit{\bm{\theta}}}(q, d) = MLP(BERT_{\mathit{cls}}([CLS] + q + [SEP] + d + [SEP])).
\end{equation}
Then, we use the pairwise cross entropy loss as $L$ in Eq.~(\ref{partial_relevance_model}) for the relevance model training:
\begin{equation}
\small
\label{bias_relevance_loss}
\mathcal{L}(R) = \frac{1}{|\mathcal{Q}|} \sum_{q \in \mathcal{Q}} \sum_{\substack{d_{i}\in \mathcal{D}^{+} \\ d_{j} \in \mathcal{D}^{ul}}} \! - \log \frac{e^{R_{\bm{\theta}}(q, d_{i})}}{e^{R_{\bm{\theta}}(q, d_{i})} + e^{R_{\bm{\theta}}(q, d_{j})}}.
\end{equation}

\textbf{Selection Model.} 
According to the construction process of labeled datasets, the pooling systems used to select documents are usually basic retrievers.
Based on this principle, we build the selection model $S$ with the same architecture as $R$.
Specifically, for the selection model $S$ (parameterized by $\bm{\phi}$), let $S_{\bm{\phi}}(q, d)$ be the estimated selection propensity for the pair $(q, d)$, which is calculated with the same method as Eq.~(\ref{relevance_score}) but with a different set of parameters.
Then we train the model $S$ by minimizing the following loss function:
\begin{equation}
\small
\label{bias_selection_loss}
\mathcal{L}(S) = \frac{1}{|\mathcal{Q}|} \sum_{q \in \mathcal{Q}} \sum_{\substack{d_{i}\in \mathcal{D}^{+} \\ d_{j} \in \mathcal{D}^{ul}}} \! - \log \frac{e^{S_{\bm{\phi}}(q, d_{i})}}{e^{S_{\bm{\phi}}(q, d_{i})} + e^{S_{\bm{\phi}}(q, d_{j})}}.
\end{equation}

\begin{algorithm}[t]
\caption{{\sc The coupled learning algorithm.}}
\setlength{\belowcaptionskip}{-0.5cm}
\label{algo}
\LinesNumbered
\KwIn{query set $\mathcal{Q}$, document set $\mathcal{D}$, the labeling $l$, $\alpha$, $\tau$}
\KwOut{relevance model $R$, selection model $S$}
    Initialize parameters $\bm{\theta}$ and $\bm{\phi}$ in models $R$ and $S$\;
    \Repeat{Convergence}{
         Sample a batch of triples $(q, d_i, d_j) \in \mathcal{B}$ from $\mathcal{D}$ with $l$ \;
        \For{$(q, d_i, d_j) \in \mathcal{B}$}{
            Estimate $w_r(q, d_i, d_j)$ with Eq.~(\ref{relevance_weight})\;
            Estimate $w_s(q, d_i, d_j)$ with Eq.~(\ref{selection_weight})\;
        }
        Compute $\mathcal{{L}^{'}}(R)$ on $\mathcal{B}$ with Eq.~(\ref{unbias_relevance_loss})\;
        Compute $\mathcal{{L}^{'}}(S)$ on $\mathcal{B}$ with Eq.~(\ref{unbias_selection_loss})\;
        $\bm{\theta}=\bm{\theta}+\alpha \cdot \frac{\partial \mathcal{{L}^{'}}(R)}{\partial \bm{\theta}}, \bm{\phi}=\bm{\phi}+\alpha \cdot \frac{\partial \mathcal{{L}^{'}}(S)}{\partial \bm{\phi}}$\;
    }
    \Return{$R$, $S$}
\end{algorithm}

\textbf{Coupled Learning Algorithm.} 
As discussed in Section~\ref{bias_analysis}, the estimation of selection propensities and relevance scores for documents is coupled with each other. 
Thus, we employ a coupled learning algorithm to train two models simultaneously and achieve bias correction learning.
During the coupled learning process, each model estimates the bias weight for the training data of the other model with their current parameters, and then injects the weight into loss functions for next update.
In this way, the relevance model $R$ and the selection model $S$ are promoted mutually.
An overview of the complete algorithm is shown in Algorithm~\ref{algo}.

We first initialize all parameters (i.e., $\bm{\theta}$ and $\bm{\phi}$) of two models.
For each pair $(q, d_{i})$, we estimate its relevance score and selection propensity with current parameters and convert them into probability distributions:
\begin{equation}
\small
\label{probability}
P(r_i^+ \!\mid\! x_i) = \frac{e^{R_{\bm{\theta}}(q, d_i)}}{\sum_{d_k \in \mathcal{D}} e^{R_{\bm{\theta}}(q, d_k)}}, \
P(s_i^+ \!\mid\! x_i) = \frac{e^{S_{\bm{\phi}}(q, d_i)}}{\sum_{d_k \in \mathcal{D}} e^{S_{\bm{\phi}}(q, d_k)}}.
\end{equation}
As shown in Eq.~(\ref{probability}), the use of the softmax function assumes that the examination probabilities on different documents in $\mathcal{D}$ will sum up to 1, which is not true in practice. This, however, does not hurt the effectiveness of models training. In fact, the predicted values of $P(r_i^+ \!\mid\! x_i)$ and $P(s_i^+ \!\mid\! x_i)$ have a minor effect on the bias correction learning as long as their relative proportions are correct. Thus, with Eq.~(\ref{unbiased_relevance_model}) \& (\ref{unbiased_selection_model}), the actual bias correction weights used in CET are:
\begin{equation}
\small
\label{relevance_weight}
w_r(q, d_i, d_j)=\frac{P(s_j^+ \mid x_j)}{P(s_i^+ \mid x_i)}=\frac{e^{S_{\bm{\phi}}(q, d_{j})/\tau}}{e^{S_{\bm{\phi}}(q, d_i)/ \tau}},
\end{equation}
\begin{equation}
\small
\label{selection_weight}
w_s(q, d_i, d_j)=\frac{P(r_j^+ \mid x_j)}{P(r_i^+ \mid x_i)}=\frac{e^{R_{\bm{\theta}}(q, d_{j})/\tau}}{e^{R_{\bm{\theta}}(q, d_i)/ \tau}},
\end{equation}
where $\tau \!>\! 0$ is a hyperparameter to control the scale of bias weights.
By injecting the bias weights into Eq.~(\ref{bias_relevance_loss}) \& (\ref{bias_selection_loss}), we can obtain the IPW loss functions:
\begin{equation}
\small
\label{unbias_relevance_loss}
\mathcal{L}^{'}(R) \! = \! \frac{1}{|\mathcal{Q}|} \sum_{q \in \mathcal{Q}} \! \sum_{\substack{d_{i}\in \mathcal{D}^{+} \\ d_{j} \in \mathcal{D}^{ul}}} \! - w_r(q, d_i, d_j) \! \cdot \log \frac{e^{R_{\bm{\theta}}(q, d_{i})}}{e^{R_{\bm{\theta}}(q, d_{i})} \! + \! e^{R_{\bm{\theta}}(q, d_{j})}}, 
\end{equation}
\begin{equation}
\small
\label{unbias_selection_loss}
\mathcal{{L}^{'}}(S) \!=\! \frac{1}{|\mathcal{Q}|} \sum_{q \in \mathcal{Q}} \! \sum_{\substack{d_i\in \mathcal{D}^{+} \\ d_j \in \mathcal{D}^{ul}}} \! - w_s(q, d_i, d_j) \! \cdot \log \frac{e^{S_{\bm{\phi}}(q, d_i)}}{e^{S_{\bm{\phi}}(q, d_i)} \!+\! e^{S_{\bm{\phi}}(q, d_j)}}.
\end{equation}
Then, parameters $\bm{\theta}$ and $\bm{\phi}$ are updated with $\mathcal{L}^{'}(R)$ and $\mathcal{L}^{'}(S)$ respectively.
This process is repeated until the algorithm converges.
After the training process, only the relevance model $R$ is retained to evaluate relevance scores for query-document pairs in the dataset.
It should be noted that although training two models simultaneously will bring higher computational cost, the evaluation process is the same inference latency as usual.

\section{Experimental Settings} 
This section presents the experimental settings, including datasets, evaluation metrics, implementation details, and baselines.

\subsection{Datasets Description}
We conduct experiments on three retrieval benchmarks to evaluate the effectiveness of CET:
\begin{itemize}[leftmargin=*]
\item \textbf{MS MARCO~\cite{nguyen2016ms}:} The \textit{passage ranking task} provides about 503k queries paired with relevant passages for training. Each query is associated with sparse relevance labels of one (or very few) passages labeled as relevant and no passages explicitly labeled as irrelevant. It also has approximately 7k queries in dev and test sets. For relevance labels creation, annotators label relevant passages from the top-10 passages retrieved by a existing retrieval system at Bing. The \textit{document ranking task} has about 367k training queries, 5k dev queries, and 5k test queries. The document that produces a relevant passage is viewed as relevant.
\item \textbf{TREC DL~\cite{craswell2020overview}:} TREC 2019 Deep Learning Track has the same training and dev set as MS MARCO, but replaces the test set with a novel set produced by TREC. It contains 43 test queries for both passage and document ranking tasks. Especially, NIST constructs separate pools for them and uses depth pooling for more complete labeling.
\item \textbf{DuReader$_{retrieval}$~\cite{qiu2022dureader_retrieval}:} DuReader$_{retrieval}$ is a Chinese dataset for passage retrieval, which contains over 90K queries and 8M passages. For training queries, annotators label relevant passages within the top-5 retrieved results by Baidu Search. To reduce false negatives in the dev and test sets, more and stronger retrievers are used for the pooling process. As a result, the average labeled relevant passages for training and dev queries are 2.57 and 4.93. 
\end{itemize}

\renewcommand{\arraystretch}{1.0}
\begin{table*}[ht]
\large
\setlength{\abovecaptionskip}{3pt}
\setlength{\belowcaptionskip}{-0.3cm}
  \centering
  \fontsize{9}{9}\selectfont
  \begin{threeparttable}
  \caption{Ranking performance on the MS MARCO and TREC DL datasets. The highest value for every column is highlighted in bold and the statistically significant (p < 0.05) improvements of our method over ANCE+BERT$_{\scriptstyle \emph{DML}}$ and ANCE+BERT$_{\scriptstyle \emph{RANCE}}$ are marked with the asterisk $\dag$ and $\ddag$ respectively.}
  \label{tab:performance_comparison}
    \begin{tabular}{lcccccccc}
    \toprule
    \multirow{3}*{Method}
    &\multicolumn{4}{c}{MS MARCO Dev}&\multicolumn{4}{c}{TREC DL Test}\cr
    \cmidrule(lr){2-5} \cmidrule(lr){6-9}
    &\multicolumn{2}{c}{Passage Ranking}&\multicolumn{2}{c}{Document Ranking}&\multicolumn{2}{c}{Passage Ranking}&\multicolumn{2}{c}{Document Ranking}\cr
    \cmidrule(lr){2-3} \cmidrule(lr){4-5} \cmidrule(lr){6-7} \cmidrule(lr){8-9}
    &MRR@10 &MRR@100 &MRR@10 &MRR@100 &NDCG@10 &NDCG@100 &NDCG@10 &NDCG@100\cr
    \toprule
    BM25 + BERT &0.3562 &0.3642 &0.3841  &0.3898 &0.6958 &0.6182 &$\bm{0.6420}$ &$\bm{0.5379}$ \cr
    ANCE + BERT &0.3403  &0.3493 &0.4091 &0.4165 &0.6971 &0.5946 &0.6268 &0.4900   \cr
    \midrule
    ANCE + BERT$_{\scriptstyle \emph{zhan}}$ &0.3486 &0.3597 &0.3903 &0.4003 &0.6905 &0.5968 &0.6115 &0.4620 \cr
    ANCE + BERT$_{\scriptstyle \emph{RocketQA}}$  &0.3476  &0.3583  &0.4083 &0.4147 &0.7018 &0.6143 &0.6257 &0.4893 \cr
    ANCE + BERT$_{\scriptstyle \emph{DML}}$ &0.3458 &0.3545  &0.4110 &0.4188 &0.7015  &0.6021 &0.6272 &0.4961   \cr
    ANCE + BERT$_{\scriptstyle \emph{RANCE}}$ &0.3513 &0.3616  &0.4118 &0.4183 &0.7032 &0.6174 &0.6270 &0.4953  \cr
    \midrule
    ANCE + BERT$_{\scriptstyle \emph{CET}}$ &$\bm{0.3638}^{\dag \ddag}$ &$\bm{0.3743}^{\dag \ddag}$ &$\bm{0.4233}^{\dag \ddag}$ &$\bm{0.4311}^{\dag \ddag}$ &$\bm{0.7243}^{\dag \ddag}$ &$\bm{0.6398}^{\dag \ddag}$ &$0.6416^{\dag \ddag}$ &$0.5289^{\dag \ddag}$ \cr
    \bottomrule
    \end{tabular}
    \end{threeparttable}
\end{table*}
\setlength{\belowcaptionskip}{-0.2cm}

\subsection{Evaluation Metrics}
We conduct the evaluation following the official settings.
For MS MARCO and TREC DL, the ranking results of top-1000 passages and top-100 documents are compared, and we use the Mean Reciprocal Rank (MRR@10 and MRR@100) for MS MARCO and Normalized Discounted Cumulative Gain (NDCG@10 and NDCG@100) for TREC DL as previous works~\cite{ma2021b, ma2021pre}. 
For the DuReader$_{retrieval}$ dataset, we report the ranking results of top-50 passages with MRR@10 and Recall@1 as officials~\cite{qiu2022dureader_retrieval}.

\subsection{Implementation Details} 
For experiments on MS MARCO and TREC DL, unless otherwise specified, we implement the relevance model and selection model with the BERT-base model~\cite{devlin2018bert} and parameters are initialized with the checkpoint released by Google\footnote{\url{https://github.com/google-research/bert}}. We adopt the popular Transformers library\footnote{\url{https://github.com/huggingface/transformers}} for implementations. 
We truncate the input sequence to a maximum of 256 tokens and 512 tokens for passage ranking and document ranking tasks respectively. 
For the document ranking task, we concatenate url, title, and body fields if they are available, and experiment with the FirstP setting~\cite{dai2019deeper}.
Negatives for passage ranking and document ranking tasks are uniformly sampled from the top-1000 and top-100 results (excluding the labeled positives) returned by given retrievers (see Section~\ref{sec:result_with_baseline} \& \ref{sec:result_with_retriever}). 

For experiments on DuReader$_{retrieval}$, we use Ernie-base~\cite{sun2019ernie} as the initialization for both relevance model and selection model.
We set the maximal length of input sequence as 384~\cite{qiu2022dureader_retrieval}.
Negatives for models training are uniformly sampled from the top-50 results (excluding the labeled positives) of retrievers (see Section~\ref{sec:result_with_baseline}).

We use the AdamW optimizer with $\beta_1 = 0.9$, $\beta_2 = 0.999$, $\epsilon = 10^{-8}$, and learning rate is $3 \times 10^{-6}$ with the rate of linear scheduling warm-up at 0.1. 
The hyperparameter $\tau$ is set to 1.0 unless noted otherwise.
For all experiments, each positive sample is paired with one negative, and no special tricks are used for training.
The batch size is set to 64 and 32 for passage and document ranking, and we run all experiments on Nvidia Tesla V100-32GB GPUs.

\subsection{Baselines}
We compare the BERT ranking model learned with CET (i.e., BERT$_{\scriptstyle \emph{CET}}$) against that learned with the following methods:
\begin{itemize}[leftmargin=*]
\item \textbf{BERT:}  It trains the naive BERT ranker without any special techniques for addressing the false negative problem.
\item \textbf{BERT$_{\scriptstyle \emph{zhan}}$:} The BERT ranker is trained with BM25 negatives and evaluated with the results of stronger retrievers. This training-evaluating inconsistency setting is employed by~\citet{zhan2020repbert} for alleviating the false negative problem in models training.
\item \textbf{BERT$_{\scriptstyle \emph{RocketQA}}$:} The denoising technique in RocketQA~\cite{ding2020rocketqa} is used for BERT rankers training. We first train a naive BERT ranker on biased data and predict relevance scores for unlabeled documents, then we randomly sample negatives with relevance scores less than $\eta$ for the denoised BERT ranker training.
\item \textbf{BERT$_{\scriptstyle \emph{RANCE}}$:} The sampling technique to remove false negatives in RANCE~\cite{prakash2021learning} is used for BERT rankers training. Following RANCE, we estimate the probability $P_{relevant}(r)$ using the dev/test set labeled with depth pooling, and then we sample negatives from the retrieval results according to $P_{relevant}(r)$. 
\item \textbf{BERT$_{\scriptstyle \emph{DML}}$:} The dynamic multi-granularity learning method~\cite{zhang2021dml} is used for BERT rankers training to address the false negatives. All parameters are set the same as in~\cite{zhang2021dml}, except that only one negative is sampled for each positive to achieve fair comparisons. 
\end{itemize}

\section{Results and Analysis} \label{results}
We empirically evaluate NRMs learned with CET to address following research questions:
\begin{itemize}[leftmargin=*]
\item \textbf{RQ1:} Can NRMs learned with CET achieve better performance as compared with other SOTA learning techniques?
\item \textbf{RQ2:} Could CET work well for training different NRMs with different retrievers for negative sampling?
\item \textbf{RQ3:} How does the hyper-parameter $\tau$ in CET affect the effectiveness of learned NRMs?
\item \textbf{RQ4:} How does CET affect the training process of NRMs on labeled datasets with pooling bias?
\end{itemize}

\renewcommand{\arraystretch}{1.0}
\begin{table}[!t]
\setlength{\abovecaptionskip}{3pt}
\setlength{\belowcaptionskip}{-0.2cm}
  \caption{Recall performance of BM25 and ANCE on the MS MARCO and TREC DL datasets.}
  \small
  \label{tab:recall}
  \setlength{\tabcolsep}{2mm}{
  \begin{tabular}{lcccc}
    \toprule
    \multirow{3}*{Model}
    &\multicolumn{2}{c}{MS MARCO Dev}  &\multicolumn{2}{c}{TREC DL Test} \cr
    \cmidrule(lr){2-3} \cmidrule(lr){4-5}
    & Passage & Document & Passage & Document  \cr
    \cmidrule(lr){2-3} \cmidrule(lr){4-5}
    & Recall@1k & Recall@100 & Recall@1k & Recall@100  \cr
    \toprule
    BM25 &0.8140 &0.7564   &0.6778 &0.3871    \cr
    ANCE  &0.9587 &0.8927   &0.6610 &0.2664   \cr
    \bottomrule
  \end{tabular}}
\end{table}
\setlength{\belowcaptionskip}{-0.5cm}

\subsection{Empirical Comparison with Baselines} \label{sec:result_with_baseline}
To answer \textbf{RQ1}, we compare CET with baseline methods on three retrieval datasets to verify the effectiveness of CET.

For MS MARCO and TREC DL, we train rankers with negatives sampled from two retrievers, i.e., BM25 and ANCE~\cite{xiong2020approximate}. We use BM25-retrieved results released by officials, and for ANCE, we use the code and checkpoint (FirstP) released by \citeauthor{xiong2020approximate}\footnote{\url{https://github.com/microsoft/ANCE}\label{code_ance}} to obtain the retrieval results. The recall performance of them are reported in Table~\ref{tab:recall}, and the ranking performance of BERT$_{\scriptstyle \emph{CET}}$ and baselines are shown in Table~\ref{tab:performance_comparison}.
From the results we can observe that:
\begin{enumerate}[leftmargin=*]
\item \textbf{BM25+BERT \textit{vs} ANCE+BERT:} For the MS MARCO dataset, recall of ANCE improves by 18\% compared with BM25 on both tasks (see Table~\ref{tab:recall}). However, the BERT ranker trained with ANCE negatives (i.e., ANCE+BERT) does not improve compared with that trained with BM25 negatives (i.e., BM25+BERT) on the passage ranking task, and improves slightly over the recall gain on the document ranking task. It indicates that without addressing the false negative, ranking models cannot benefit from stronger retrievers for negative sampling. Besides, it is worth mention that on the document ranking task of TREC DL, Recall@100 of ANCE is 31\% lower than BM25 (see Table~\ref{tab:recall}), causing ANCE+BERT performs significantly worse than BM25+BERT. 
\item \textbf{ANCE+BERT \textit{vs} ANCE+BERT$_{\scriptstyle \star}$:}
For BERT rankers trained with special techniques to solve the false negative issue, we collectively call them BERT$_{\scriptstyle \star}$ here for simplicity.
Among them, ANCE+BERT$_{\scriptstyle \emph{zhan}}$ performs better than ANCE+BERT on the MS MARCO passage ranking task. However, this method generally fails on the other three datasets.
With the denoising technique, BERT$_{\scriptstyle \emph{RocketQA}}$ can alleviate false negatives on passage ranking task, but it does not work for document ranking. 
ANCE+BERT$_{\scriptstyle \emph{DML}}$ could slightly improve the performance over ANCE+BERT, consisting with the result in~\cite{zhang2021dml}.
BERT$_{\scriptstyle \emph{RANCE}}$ and BERT$_{\scriptstyle \emph{CET}}$ both achieve improvements over naive BERT rankers on all the passage ranking and document ranking tasks.
\item \textbf{BERT$_{\scriptstyle \emph{DML}}$ \textit{vs} BERT$_{\scriptstyle \emph{RANCE}}$ \textit{vs} BERT$_{\scriptstyle \emph{CET}}$:} 
To further analyze the result, ANCE+BERT$_{\scriptstyle \emph{DML}}$ and ANCE+BERT$_{\scriptstyle \emph{RANCE}}$ do not show advantages over ANCE+BERT on the document ranking task. 
Different from them, ANCE+BERT$_{\scriptstyle \emph{CET}}$ improves significantly than ANCE+BERT on all tasks. 
Overall, among three methods, CET performs best and shows stable superiority against all baseline methods.
Besides, it is worth noting that the gain of CET is more significant on the passage ranking task.
For example, for the MS MARCO dataset, ANCE+BERT$_{\scriptstyle \emph{CET}}$ improves about 6.9\% and 3.5\% on MRR@10 over ANCE+BERT for the passage ranking and document ranking tasks respectively. 
\end{enumerate}

For the DuReader$_{retrieval}$, we train rankers based on the retrieved results by BM25 and a strong retriever (called DE in~\cite{qiu2022dureader_retrieval}). We directly use the DE-retrieved results released by officials, and we get BM25-retrieved results with Anserini\footnote{\url{http://anserini.io/}} since the official result is unavailable. 
The retrieval performance in terms of Recall@50 of BM25 and DE is 0.6635 and 0.9115, respectively.
The ranking performance is shown in Table~\ref{tab:ranking_dureader}.
From the results, we can find that: 
\begin{enumerate}[leftmargin=*]
\item \textbf{BM25+BERT \textit{vs} DE+BERT:} With stronger retrievers to obtain hard negatives for rankers training (e.g, Recall@50 of DE improves 37\% over BM25), DE+BERT could improve 33\% on MRR@10 over BM25+BERT. We speculate that it is because the false negative problem in DuReader$_{retrieval}$ is slighter than that in MS MARCO dataset. This could be further proved by comparing the number of labeled relevant passages before and after depth pooling for two datasets~\cite{craswell2020overview, qiu2022dureader_retrieval}.
\item \textbf{DE+BERT \textit{vs} DE+BERT$_{\scriptstyle \star}$:} In spite of the slighter bias problem in DuReader$_{retrieval}$, BERT rankers learned with special techniques to solve the false negative could further improve the ranking performance than the naive BERT ranker. Especially, compared with all baselines, the performance improvement of CET is the most significant, where the gain on MRR@10 and Recall@1 are 2.2\% and 4.4\% respectively. 
\end{enumerate}

In summary, BERT rankers learned with CET perform better on ranking effectiveness over baselines. Empirically, our method is able to correct pooling bias in training NRMs on labeled datasets.

\renewcommand{\arraystretch}{1.0}
\begin{table}[!t]
\setlength{\abovecaptionskip}{3pt}
\setlength{\belowcaptionskip}{-0.2cm}
  \caption{Ranking performance on DuReader$_{retrieval}$. The highest value for each column is highlighted in bold and the statistically significant (p < 0.05) improvements of our method over DE+BERT$_{\scriptstyle \emph{RocketQA}}$ are marked with $\dag$.}
  \small
  \label{tab:ranking_dureader}
  \setlength{\tabcolsep}{3.5mm}{
  \begin{tabular}{lccc}
    \toprule
    Model &MRR@10 &Recall@1   \cr  
    \toprule
    BM25 + BERT &0.5391 &0.4675    \cr  
    DE + BERT &0.7168 &0.6210    \cr  
    \midrule
    DE + BERT$_{\scriptstyle \emph{DML}}$ &0.7171 &0.6230  \cr
    DE + BERT$_{\scriptstyle \emph{RANCE}}$ &0.7196 &0.6265  \cr
    DE + BERT$_{\scriptstyle \emph{RocketQA}}$ &0.7223 &0.6300  \cr
    \midrule
    DE + BERT$_{\scriptstyle \emph{CET}}$ &$\bm{0.7323}^{\dag}$ &$\bm{0.6485}^{\dag}$   \cr
    \bottomrule
  \end{tabular}}
\end{table}
\setlength{\belowcaptionskip}{-0.2cm}

\renewcommand{\arraystretch}{1.0}
\begin{table}[!t]
\setlength{\abovecaptionskip}{3pt}
\setlength{\belowcaptionskip}{-0.2cm}
  \caption{MRR@100 of different rankers learned with or without CET on MS MARCO document ranking task.}
  \small
  \label{tab:ranking}
  \setlength{\tabcolsep}{2mm}{
  \begin{tabular}{ccccc}
    \toprule
    Model &BERT &PROP &CEDR-KNRM &PARADE-Max \cr
    \toprule
    w/o CET  &0.4165 &0.4293 &0.4317 &0.4278 \cr
    w/ CET &0.4311 &0.4400 &0.4403 &0.4373 \cr
    \bottomrule
  \end{tabular}}
\end{table}
\setlength{\belowcaptionskip}{-0.5cm}

\begin{figure*}[!t]
\setlength{\abovecaptionskip}{2pt}
\centering
\includegraphics[scale=0.4]{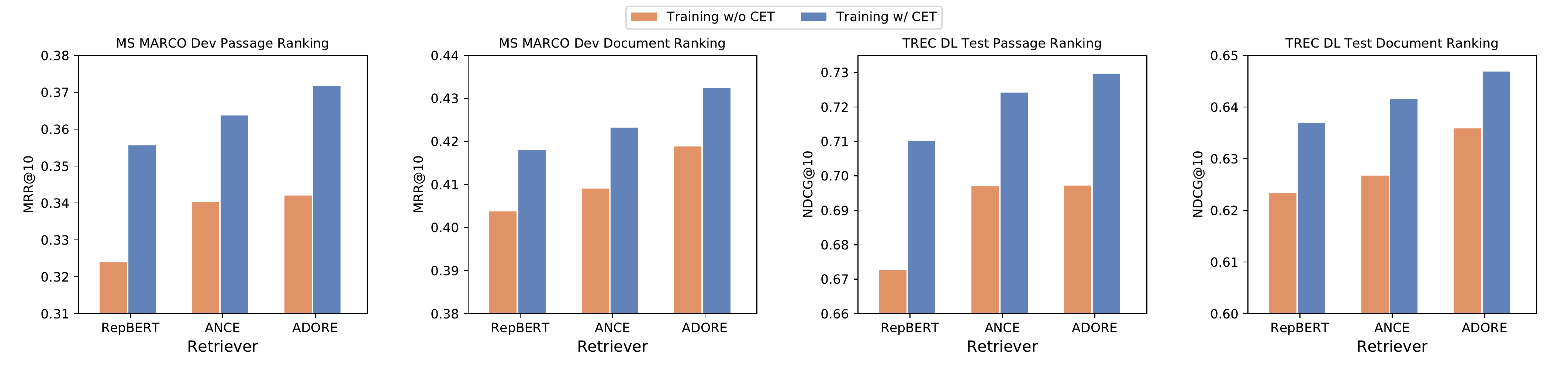}
\caption{Performance of BERT rankers learned with/without CET along with three different retrievers to sample negatives.} 
\label{different_retriever}                                  
\end{figure*}

\subsection{Results with Different Rankers}
To answer \textbf{RQ2}, we conduct experiments to investigate whether CET works well with more advanced ranking models. Specifically, we take three NRMs, i.e., PROP~\cite{ma2021prop}, CEDR~\cite{macavaney2019cedr}, and PARADE~\cite{li2020parade} as the ranker, and train them with negatives sampled from ANCE retrieved results on the MS MARCO document ranking task.
For PROP, we use the checkpoint released by authors\footnote{\url{https://github.com/Albert-Ma/PROP}} as the initialization of the ranking model.
For CEDR, we choose the best variant, CEDR-KNRM\footnote{\url{https://github.com/Georgetown-IR-Lab/cedr}}, to conduct experiments. 
For PARADE, we use the BERT ranker fine-tuned on MS MARCO passage ranking task (released by authors\footnote{\url{https://github.com/canjiali/PARADE}}) to initialize PARADE-Max’s PLM component.

Table~\ref{tab:ranking} reports the performance comparison of four rankers learned with or without CET. From the results, we can observe that even learning without CET, three advanced ranking models all perform better than the BERT ranker, which verifies the effectiveness of these techniques.
On the other hand, it is obvious that these models could achieve better ranking performance after equipped with CET. It indicates that CET is effective for correcting pooling bias in training different NRMs.

\subsection{Results with Different Retrievers} \label{sec:result_with_retriever}
To answer \textbf{RQ2}, we further investigate whether CET works well with negatives sampled from different retrievers. For this purpose, we take two other strong retrievers, i.e., RepBERT~\cite{zhan2020repbert} and ADORE~\cite{zhan2021optimizing}, along with ANCE for experiments.
For RepBERT, we train it on the document ranking task following~\cite{zhan2020repbert} and adopt their open-source codes\footnote{\url{https://github.com/jingtaozhan/RepBERT-Index}\label{code_repbert}}. For other experiments, we use the code and checkpoint released by authors\textsuperscript{\ref{code_ance},\ref{code_repbert},}\footnote{\url{https://github.com/jingtaozhan/DRhard}} to obtain the retrieval results. 

Here, we report the ranking performances on MS MARCO and TREC DL with the above three retrievers for negative sampling, and results are depicted in Fig.~\ref{different_retriever}.
We can observe that with different retrievers for negative sampling, BERT rankers learned with CET all improve significantly than that without CET. It indicates that CET works well for addressing false negatives from different retrievers.
Furthermore, comparing results on two tasks, the gain of CET is more obvious on the passage ranking task, e.g., averagely 8.5\% \textit{vs} 3.5\% on passage ranking and document ranking for MS MARCO. 

\begin{figure}[!t]
\setlength{\abovecaptionskip}{3pt}
\includegraphics[scale=0.55]{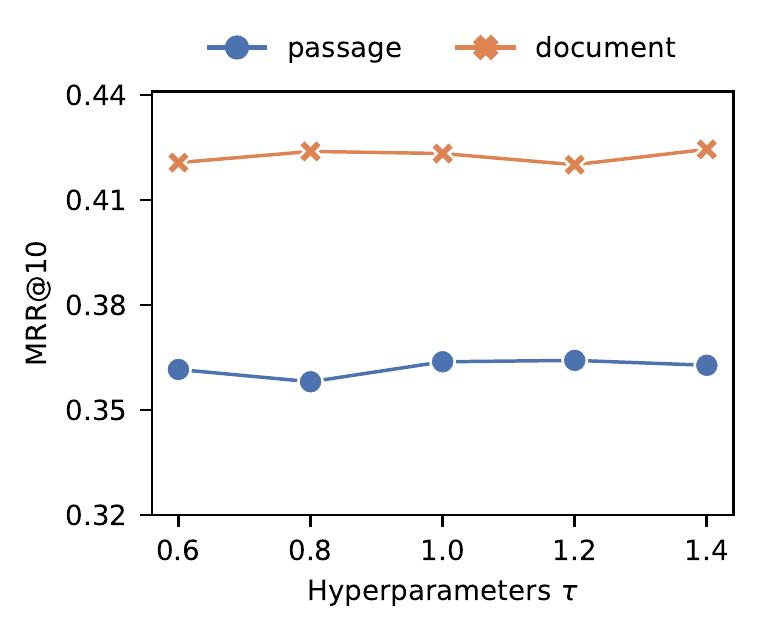}
\hspace{1mm}
\includegraphics[scale=0.55]{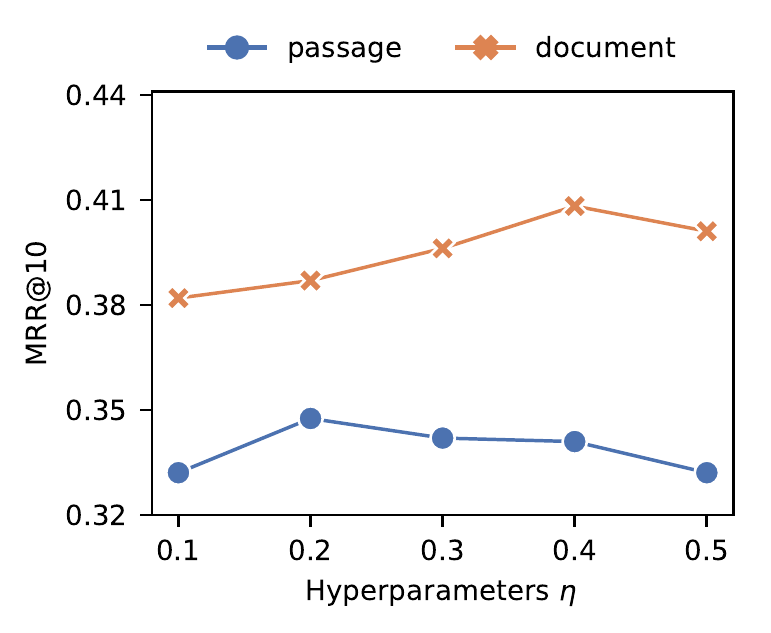}
\caption{Performance of BERT$_{\scriptstyle \emph{CET}}$ (left) and BERT$_{\scriptstyle \emph{RocketQA}}$ (right) w.r.t their hyperparameters on MS MARCO dataset.}
\label{temperature_coefficient}
\end{figure}

\subsection{Parameter Sensitivity Analysis}
To answer \textbf{RQ3}, we evaluate ANCE+BERT$_{\scriptstyle \emph{CET}}$ with different values of $\tau$ to investigate its impact on models effectiveness.

As shown in Fig.~\ref{temperature_coefficient}, we report MRR@10 of ANCE+BERT$_{\scriptstyle \emph{CET}}$ and ANCE+BERT$_{\scriptstyle \emph{RocketQA}}$ with different choices for $\tau$ and $\eta$ on both tasks of MS MARCO.
We can see that the impact of $\tau$ is slight on the performance of BERT rankers learned with CET, while the denoising technique in RocketQA~\cite{ding2020rocketqa} is highly sensitive w.r.t its hyperparameters $\eta$ (e.g., the performance variance of ANCE+BERT$_{\scriptstyle \emph{RocketQA}}$ is about 10 times larger than that of ANCE+BERT$_{\scriptstyle \emph{CET}}$ on the passage ranking task).
We suspect that it is because the value of $\eta$ directly determines the extent of denoising. 
If it is set too large, false negatives cannot be fully removed, and conversely, hard negatives would be filtered at the same time. Both of cases would hurt rankers training.
Especially, compared with the result in Table~\ref{tab:performance_comparison}, we can find that if $\eta$ is set unreasonably, the BERT ranker trained with the denoising technique would be worse than that without denoising.
By comparison, CET is much stable and robust in performance.

\subsection{Probing Analysis}
\begin{figure}[!t]
\setlength{\abovecaptionskip}{3pt}
\centering
\includegraphics[scale=0.4]{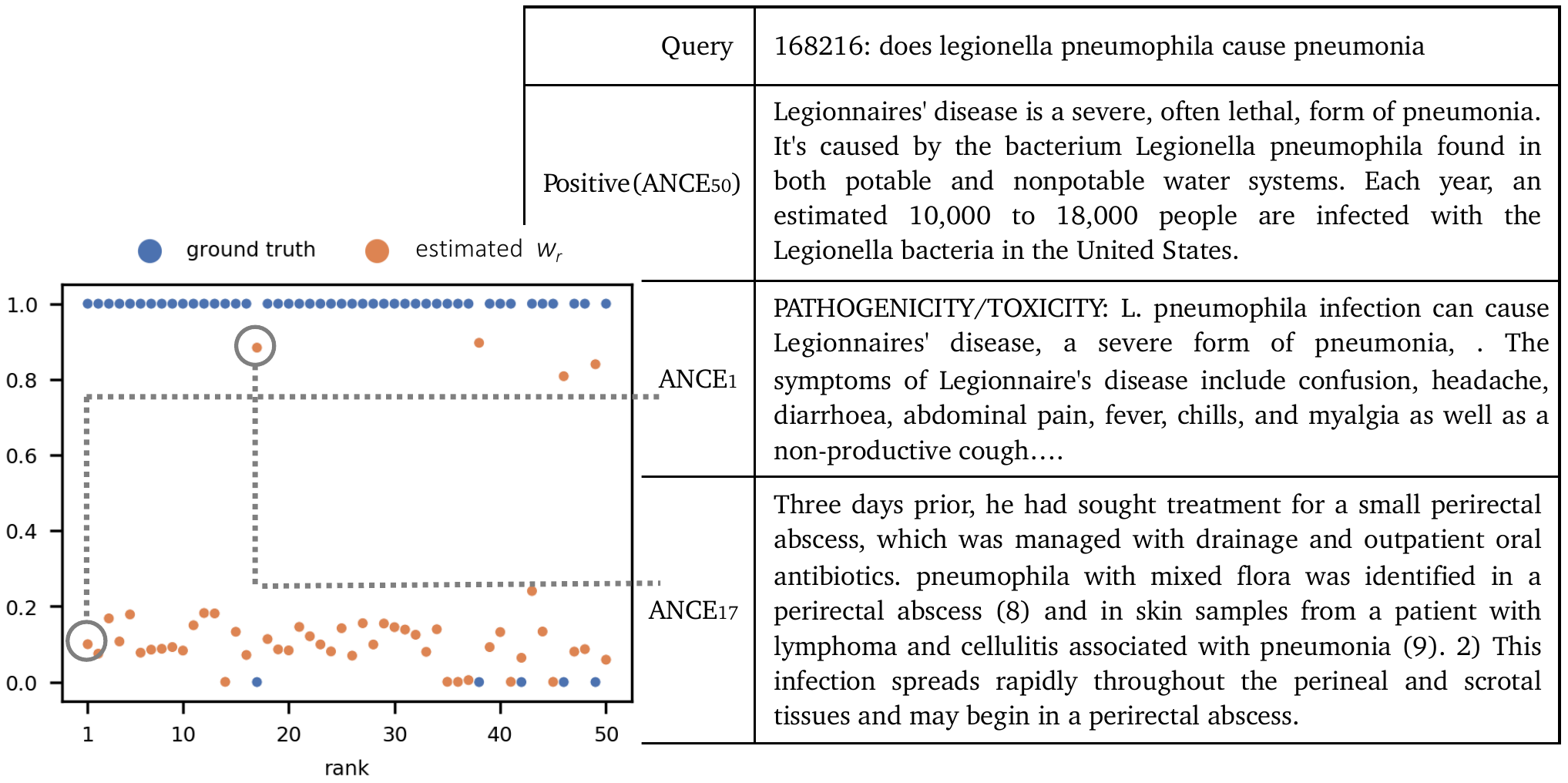}
\caption{A case to show the ground truth relevance and estimated $w_r$ of top-50 documents retrieved by ANCE.}
\label{case_study}
\end{figure}

To answer \textbf{RQ4}, we investigate how NRMs perform in distinguishing hard negatives and false negatives during the training process.
Specifically, we select a query $q$ from the passage ranking task of TREC DL Test set. Then, we take the selection model checkpoint (Step 150k) to estimate $w_r(q, d_i, d_j)$ for the top-50 documents retrieved by ANCE as $d_j$, paired with one of labeled relevant documents (ranked at position 51 by ANCE) as $d_i$. We show the distribution of the ground truth relevance and estimated $w_r$ in Fig.~\ref{case_study}.

With more complete labeled ground truth in the TREC DL test set, we can see that 45 of the top-50 retrieval results by ANCE are relevant documents for Query \#168216. It further indicates that there are riddled false negatives in the training set, where usually one (at most four) passages are labeled as relevant.
Besides, it shows a negative correlation between the ground truth and estimated $w_r$ for the top-50 documents. 
That is, during the coupled learning process, the selection model would estimate lower $w_r$ if $d_j$ is an unlabeled relevant document.
It implies that the relevance model learned with the IPW loss function (i.e., Eq.~(\ref{unbias_relevance_loss})) could relax the penalty to false negatives. This intuitively displays how CET learns to adaptively distinguish false negatives and hard negatives and achieve bias correction learning.

\section{Conclusion}
In this work, we formulate the false negative issue in training NRMs as learning from labeled datasets with pooling bias.
To address this problem, we follow the inverse propensity weighting learning framework and propose a Coupled Estimation Technique.
CET learns both a relevance model and a selection model simultaneously with a coupled learning algorithm on the biased dataset.
During the training process, the ranking model could adaptively distinguish hard negatives and false negatives with the selection propensities estimated by the selection model and achieve bias correction learning.
Empirical results on three retrieval benchmarks show that NRMs learned with CET achieve significant gains on ranking effectiveness against baseline methods.
In the future, we would like to extend this method to the listwise setting, where multiple negative samples are considered simultaneously to estimate the bias weights. 
Moreover, we would also try to apply this method in the first-stage retrieval to address the pooling bias problem.

\begin{acks}
This work was funded by the National Natural Science Foundation of China (NSFC) under Grants No. 61902381 and 62006218, the Youth Innovation Promotion Association CAS under Grants No. 20144310, and 2021100, the Young Elite Scientist Sponsorship Program by CAST under Grants No. YESS20200121, and the Lenovo-CAS Joint Lab Youth Scientist Project.
\end{acks}

\bibliographystyle{ACM-Reference-Format}
\bibliography{paper}


\begin{thebibliography}{52}


\ifx \showCODEN    \undefined \def \showCODEN     #1{\unskip}     \fi
\ifx \showDOI      \undefined \def \showDOI       #1{#1}\fi
\ifx \showISBNx    \undefined \def \showISBNx     #1{\unskip}     \fi
\ifx \showISBNxiii \undefined \def \showISBNxiii  #1{\unskip}     \fi
\ifx \showISSN     \undefined \def \showISSN      #1{\unskip}     \fi
\ifx \showLCCN     \undefined \def \showLCCN      #1{\unskip}     \fi
\ifx \shownote     \undefined \def \shownote      #1{#1}          \fi
\ifx \showarticletitle \undefined \def \showarticletitle #1{#1}   \fi
\ifx \showURL      \undefined \def \showURL       {\relax}        \fi
\providecommand\bibfield[2]{#2}
\providecommand\bibinfo[2]{#2}
\providecommand\natexlab[1]{#1}
\providecommand\showeprint[2][]{arXiv:#2}

\bibitem[\protect\citeauthoryear{Ai, Bi, Luo, Guo, and Croft}{Ai
  et~al\mbox{.}}{2018a}]%
        {ai2018unbiased}
\bibfield{author}{\bibinfo{person}{Qingyao Ai}, \bibinfo{person}{Keping Bi},
  \bibinfo{person}{Cheng Luo}, \bibinfo{person}{Jiafeng Guo}, {and}
  \bibinfo{person}{W~Bruce Croft}.} \bibinfo{year}{2018}\natexlab{a}.
\newblock \showarticletitle{Unbiased learning to rank with unbiased propensity
  estimation}. In \bibinfo{booktitle}{\emph{The 41st International ACM SIGIR
  Conference on Research \& Development in Information Retrieval}}.
  \bibinfo{pages}{385--394}.
\newblock


\bibitem[\protect\citeauthoryear{Ai, Mao, Liu, and Croft}{Ai
  et~al\mbox{.}}{2018b}]%
        {ai2018ULTR}
\bibfield{author}{\bibinfo{person}{Qingyao Ai}, \bibinfo{person}{Jiaxin Mao},
  \bibinfo{person}{Yiqun Liu}, {and} \bibinfo{person}{W~Bruce Croft}.}
  \bibinfo{year}{2018}\natexlab{b}.
\newblock \showarticletitle{Unbiased learning to rank: Theory and practice}. In
  \bibinfo{booktitle}{\emph{Proceedings of the 27th ACM CIKM}}.
  \bibinfo{pages}{2305--2306}.
\newblock


\bibitem[\protect\citeauthoryear{Arabzadeh, Vtyurina, Yan, and
  Clarke}{Arabzadeh et~al\mbox{.}}{2021}]%
        {arabzadeh2021shallow}
\bibfield{author}{\bibinfo{person}{Negar Arabzadeh}, \bibinfo{person}{Alexandra
  Vtyurina}, \bibinfo{person}{Xinyi Yan}, {and} \bibinfo{person}{Charles~LA
  Clarke}.} \bibinfo{year}{2021}\natexlab{}.
\newblock \showarticletitle{Shallow pooling for sparse labels}.
\newblock \bibinfo{journal}{\emph{arXiv preprint arXiv:2109.00062}}
  (\bibinfo{year}{2021}).
\newblock


\bibitem[\protect\citeauthoryear{Arabzadeh, Vtyurina, Yan, and
  Clarke}{Arabzadeh et~al\mbox{.}}{2022}]%
        {arabzadeh2022shallow}
\bibfield{author}{\bibinfo{person}{Negar Arabzadeh}, \bibinfo{person}{Alexandra
  Vtyurina}, \bibinfo{person}{Xinyi Yan}, {and} \bibinfo{person}{Charles~LA
  Clarke}.} \bibinfo{year}{2022}\natexlab{}.
\newblock \showarticletitle{Shallow pooling for sparse labels}.
\newblock \bibinfo{journal}{\emph{Information Retrieval Journal}}
  (\bibinfo{year}{2022}), \bibinfo{pages}{1--21}.
\newblock


\bibitem[\protect\citeauthoryear{Buckley, Dimmick, Soboroff, and
  Voorhees}{Buckley et~al\mbox{.}}{2006}]%
        {buckley2006bias}
\bibfield{author}{\bibinfo{person}{Chris Buckley}, \bibinfo{person}{Darrin
  Dimmick}, \bibinfo{person}{Ian Soboroff}, {and} \bibinfo{person}{Ellen
  Voorhees}.} \bibinfo{year}{2006}\natexlab{}.
\newblock \showarticletitle{Bias and the limits of pooling}. In
  \bibinfo{booktitle}{\emph{Proceedings of the 29th annual international ACM
  SIGIR}}. \bibinfo{pages}{619--620}.
\newblock


\bibitem[\protect\citeauthoryear{Burges}{Burges}{2010}]%
        {burges2010ranknet}
\bibfield{author}{\bibinfo{person}{Christopher~JC Burges}.}
  \bibinfo{year}{2010}\natexlab{}.
\newblock \showarticletitle{From ranknet to lambdarank to lambdamart: An
  overview}.
\newblock \bibinfo{journal}{\emph{Learning}} \bibinfo{volume}{11},
  \bibinfo{number}{23-581} (\bibinfo{year}{2010}), \bibinfo{pages}{81}.
\newblock


\bibitem[\protect\citeauthoryear{Clarke, Craswell, and Soboroff}{Clarke
  et~al\mbox{.}}{2009}]%
        {clarke2009overview}
\bibfield{author}{\bibinfo{person}{Charles~L Clarke}, \bibinfo{person}{Nick
  Craswell}, {and} \bibinfo{person}{Ian Soboroff}.}
  \bibinfo{year}{2009}\natexlab{}.
\newblock \bibinfo{booktitle}{\emph{Overview of the trec 2009 web track}}.
\newblock \bibinfo{type}{{T}echnical {R}eport}. \bibinfo{institution}{WATERLOO
  UNIV (ONTARIO)}.
\newblock


\bibitem[\protect\citeauthoryear{Clarke, Craswell, Soboroff,
  et~al\mbox{.}}{Clarke et~al\mbox{.}}{2004}]%
        {clarke2004overview}
\bibfield{author}{\bibinfo{person}{Charles~LA Clarke}, \bibinfo{person}{Nick
  Craswell}, \bibinfo{person}{Ian Soboroff}, {et~al\mbox{.}}}
  \bibinfo{year}{2004}\natexlab{}.
\newblock \showarticletitle{Overview of the TREC 2004 Terabyte Track.}. In
  \bibinfo{booktitle}{\emph{TREC}}, Vol.~\bibinfo{volume}{4}.
  \bibinfo{pages}{74}.
\newblock


\bibitem[\protect\citeauthoryear{Craswell, Mitra, Yilmaz, Campos, and
  Voorhees}{Craswell et~al\mbox{.}}{2020}]%
        {craswell2020overview}
\bibfield{author}{\bibinfo{person}{Nick Craswell}, \bibinfo{person}{Bhaskar
  Mitra}, \bibinfo{person}{Emine Yilmaz}, \bibinfo{person}{Daniel Campos},
  {and} \bibinfo{person}{Ellen~M Voorhees}.} \bibinfo{year}{2020}\natexlab{}.
\newblock \showarticletitle{Overview of the trec 2019 deep learning track}.
\newblock \bibinfo{journal}{\emph{arXiv preprint arXiv:2003.07820}}
  (\bibinfo{year}{2020}).
\newblock


\bibitem[\protect\citeauthoryear{Dai and Callan}{Dai and Callan}{2019}]%
        {dai2019deeper}
\bibfield{author}{\bibinfo{person}{Zhuyun Dai} {and} \bibinfo{person}{Jamie
  Callan}.} \bibinfo{year}{2019}\natexlab{}.
\newblock \showarticletitle{Deeper text understanding for IR with contextual
  neural language modeling}. In \bibinfo{booktitle}{\emph{Proceedings of the
  42nd International ACM SIGIR}}. \bibinfo{pages}{985--988}.
\newblock


\bibitem[\protect\citeauthoryear{Devlin, Chang, Lee, and Toutanova}{Devlin
  et~al\mbox{.}}{2018}]%
        {devlin2018bert}
\bibfield{author}{\bibinfo{person}{Jacob Devlin}, \bibinfo{person}{Ming-Wei
  Chang}, \bibinfo{person}{Kenton Lee}, {and} \bibinfo{person}{Kristina
  Toutanova}.} \bibinfo{year}{2018}\natexlab{}.
\newblock \showarticletitle{Bert: Pre-training of deep bidirectional
  transformers for language understanding}.
\newblock \bibinfo{journal}{\emph{arXiv preprint arXiv:1810.04805}}
  (\bibinfo{year}{2018}).
\newblock


\bibitem[\protect\citeauthoryear{Ding, Quan, Yao, Li, and Jin}{Ding
  et~al\mbox{.}}{2020b}]%
        {ding2020simplify}
\bibfield{author}{\bibinfo{person}{Jingtao Ding}, \bibinfo{person}{Yuhan Quan},
  \bibinfo{person}{Quanming Yao}, \bibinfo{person}{Yong Li}, {and}
  \bibinfo{person}{Depeng Jin}.} \bibinfo{year}{2020}\natexlab{b}.
\newblock \showarticletitle{Simplify and Robustify Negative Sampling for
  Implicit Collaborative Filtering}.
\newblock \bibinfo{journal}{\emph{arXiv preprint arXiv:2009.03376}}
  (\bibinfo{year}{2020}).
\newblock


\bibitem[\protect\citeauthoryear{Ding, Liu, Liu, Ren, Zhao, Dong, Wu, and
  Wang}{Ding et~al\mbox{.}}{2020a}]%
        {ding2020rocketqa}
\bibfield{author}{\bibinfo{person}{Yingqi Qu~Yuchen Ding},
  \bibinfo{person}{Jing Liu}, \bibinfo{person}{Kai Liu},
  \bibinfo{person}{Ruiyang Ren}, \bibinfo{person}{Xin Zhao},
  \bibinfo{person}{Daxiang Dong}, \bibinfo{person}{Hua Wu}, {and}
  \bibinfo{person}{Haifeng Wang}.} \bibinfo{year}{2020}\natexlab{a}.
\newblock \showarticletitle{RocketQA: An Optimized Training Approach to Dense
  Passage Retrieval for Open-Domain Question Answering}.
\newblock \bibinfo{journal}{\emph{arXiv preprint arXiv:2010.08191}}
  (\bibinfo{year}{2020}).
\newblock


\bibitem[\protect\citeauthoryear{Gao, Dai, and Callan}{Gao
  et~al\mbox{.}}{2021}]%
        {gao2021rethink}
\bibfield{author}{\bibinfo{person}{Luyu Gao}, \bibinfo{person}{Zhuyun Dai},
  {and} \bibinfo{person}{Jamie Callan}.} \bibinfo{year}{2021}\natexlab{}.
\newblock \showarticletitle{Rethink training of BERT rerankers in multi-stage
  retrieval pipeline}.
\newblock \bibinfo{journal}{\emph{arXiv preprint arXiv:2101.08751}}
  (\bibinfo{year}{2021}).
\newblock


\bibitem[\protect\citeauthoryear{Guo, Cai, Fan, Sun, Zhang, and Cheng}{Guo
  et~al\mbox{.}}{2022}]%
        {guo2022semantic}
\bibfield{author}{\bibinfo{person}{Jiafeng Guo}, \bibinfo{person}{Yinqiong
  Cai}, \bibinfo{person}{Yixing Fan}, \bibinfo{person}{Fei Sun},
  \bibinfo{person}{Ruqing Zhang}, {and} \bibinfo{person}{Xueqi Cheng}.}
  \bibinfo{year}{2022}\natexlab{}.
\newblock \showarticletitle{Semantic models for the first-stage retrieval: A
  comprehensive review}.
\newblock \bibinfo{journal}{\emph{ACM Transactions on Information Systems
  (TOIS)}} \bibinfo{volume}{40}, \bibinfo{number}{4} (\bibinfo{year}{2022}),
  \bibinfo{pages}{1--42}.
\newblock


\bibitem[\protect\citeauthoryear{Guo, Fan, Pang, Yang, Ai, Zamani, Wu, Croft,
  and Cheng}{Guo et~al\mbox{.}}{2020}]%
        {guo2020deep}
\bibfield{author}{\bibinfo{person}{Jiafeng Guo}, \bibinfo{person}{Yixing Fan},
  \bibinfo{person}{Liang Pang}, \bibinfo{person}{Liu Yang},
  \bibinfo{person}{Qingyao Ai}, \bibinfo{person}{Hamed Zamani},
  \bibinfo{person}{Chen Wu}, \bibinfo{person}{W~Bruce Croft}, {and}
  \bibinfo{person}{Xueqi Cheng}.} \bibinfo{year}{2020}\natexlab{}.
\newblock \showarticletitle{A deep look into neural ranking models for
  information retrieval}.
\newblock \bibinfo{journal}{\emph{Information Processing \& Management}}
  \bibinfo{volume}{57}, \bibinfo{number}{6} (\bibinfo{year}{2020}),
  \bibinfo{pages}{102067}.
\newblock


\bibitem[\protect\citeauthoryear{Hu, Wang, Peng, and Li}{Hu
  et~al\mbox{.}}{2019}]%
        {Unbiased2019Hu}
\bibfield{author}{\bibinfo{person}{Ziniu Hu}, \bibinfo{person}{Yang Wang},
  \bibinfo{person}{Qu Peng}, {and} \bibinfo{person}{Hang Li}.}
  \bibinfo{year}{2019}\natexlab{}.
\newblock \showarticletitle{Unbiased lambdamart: an unbiased pairwise
  learning-to-rank algorithm}. In \bibinfo{booktitle}{\emph{The WWW
  Conference}}. \bibinfo{pages}{2830--2836}.
\newblock


\bibitem[\protect\citeauthoryear{Huang, He, Gao, Deng, Acero, and Heck}{Huang
  et~al\mbox{.}}{2013}]%
        {huang2013learning}
\bibfield{author}{\bibinfo{person}{Po-Sen Huang}, \bibinfo{person}{Xiaodong
  He}, \bibinfo{person}{Jianfeng Gao}, \bibinfo{person}{Li Deng},
  \bibinfo{person}{Alex Acero}, {and} \bibinfo{person}{Larry Heck}.}
  \bibinfo{year}{2013}\natexlab{}.
\newblock \showarticletitle{Learning deep structured semantic models for web
  search using clickthrough data}. In \bibinfo{booktitle}{\emph{Proceedings of
  the 22nd ACM CIKM}}. \bibinfo{pages}{2333--2338}.
\newblock


\bibitem[\protect\citeauthoryear{Joachims, Swaminathan, and Schnabel}{Joachims
  et~al\mbox{.}}{2017}]%
        {joachims2017unbiased}
\bibfield{author}{\bibinfo{person}{Thorsten Joachims}, \bibinfo{person}{Adith
  Swaminathan}, {and} \bibinfo{person}{Tobias Schnabel}.}
  \bibinfo{year}{2017}\natexlab{}.
\newblock \showarticletitle{Unbiased learning-to-rank with biased feedback}. In
  \bibinfo{booktitle}{\emph{Proceedings of the Tenth ACM WSDM}}.
  \bibinfo{pages}{781--789}.
\newblock


\bibitem[\protect\citeauthoryear{Karpukhin, O{\u{g}}uz, Min, Wu, Edunov, Chen,
  and Yih}{Karpukhin et~al\mbox{.}}{2020}]%
        {karpukhin2020dense}
\bibfield{author}{\bibinfo{person}{Vladimir Karpukhin}, \bibinfo{person}{Barlas
  O{\u{g}}uz}, \bibinfo{person}{Sewon Min}, \bibinfo{person}{Ledell Wu},
  \bibinfo{person}{Sergey Edunov}, \bibinfo{person}{Danqi Chen}, {and}
  \bibinfo{person}{Wen-tau Yih}.} \bibinfo{year}{2020}\natexlab{}.
\newblock \showarticletitle{Dense Passage Retrieval for Open-Domain Question
  Answering}.
\newblock \bibinfo{journal}{\emph{arXiv preprint arXiv:2004.04906}}
  (\bibinfo{year}{2020}).
\newblock


\bibitem[\protect\citeauthoryear{Kwiatkowski, Palomaki, Redfield, Collins,
  Parikh, Alberti, Epstein, Polosukhin, Devlin, Lee, et~al\mbox{.}}{Kwiatkowski
  et~al\mbox{.}}{2019}]%
        {kwiatkowski2019natural}
\bibfield{author}{\bibinfo{person}{Tom Kwiatkowski},
  \bibinfo{person}{Jennimaria Palomaki}, \bibinfo{person}{Olivia Redfield},
  \bibinfo{person}{Michael Collins}, \bibinfo{person}{Ankur Parikh},
  \bibinfo{person}{Chris Alberti}, \bibinfo{person}{Danielle Epstein},
  \bibinfo{person}{Illia Polosukhin}, \bibinfo{person}{Jacob Devlin},
  \bibinfo{person}{Kenton Lee}, {et~al\mbox{.}}}
  \bibinfo{year}{2019}\natexlab{}.
\newblock \showarticletitle{Natural questions: a benchmark for question
  answering research}.
\newblock \bibinfo{journal}{\emph{Transactions of the Association for
  Computational Linguistics}}  \bibinfo{volume}{7} (\bibinfo{year}{2019}),
  \bibinfo{pages}{453--466}.
\newblock


\bibitem[\protect\citeauthoryear{Lakshman, Teo, Chu, Nigam, Patni, Maknikar,
  and Vishwanathan}{Lakshman et~al\mbox{.}}{2021}]%
        {lakshman2021embracing}
\bibfield{author}{\bibinfo{person}{Vihan Lakshman}, \bibinfo{person}{Choon~Hui
  Teo}, \bibinfo{person}{Xiaowen Chu}, \bibinfo{person}{Priyanka Nigam},
  \bibinfo{person}{Abhinandan Patni}, \bibinfo{person}{Pooja Maknikar}, {and}
  \bibinfo{person}{SVN Vishwanathan}.} \bibinfo{year}{2021}\natexlab{}.
\newblock \showarticletitle{Embracing Structure in Data for Billion-Scale
  Semantic Product Search}.
\newblock \bibinfo{journal}{\emph{arXiv preprint arXiv:2110.06125}}
  (\bibinfo{year}{2021}).
\newblock


\bibitem[\protect\citeauthoryear{Li, Yates, MacAvaney, He, and Sun}{Li
  et~al\mbox{.}}{2020}]%
        {li2020parade}
\bibfield{author}{\bibinfo{person}{Canjia Li}, \bibinfo{person}{Andrew Yates},
  \bibinfo{person}{Sean MacAvaney}, \bibinfo{person}{Ben He}, {and}
  \bibinfo{person}{Yingfei Sun}.} \bibinfo{year}{2020}\natexlab{}.
\newblock \showarticletitle{PARADE: Passage representation aggregation for
  document reranking}.
\newblock \bibinfo{journal}{\emph{arXiv preprint arXiv:2008.09093}}
  (\bibinfo{year}{2020}).
\newblock


\bibitem[\protect\citeauthoryear{Ma, Guo, Zhang, Fan, Ji, and Cheng}{Ma
  et~al\mbox{.}}{2021b}]%
        {ma2021prop}
\bibfield{author}{\bibinfo{person}{Xinyu Ma}, \bibinfo{person}{Jiafeng Guo},
  \bibinfo{person}{Ruqing Zhang}, \bibinfo{person}{Yixing Fan},
  \bibinfo{person}{Xiang Ji}, {and} \bibinfo{person}{Xueqi Cheng}.}
  \bibinfo{year}{2021}\natexlab{b}.
\newblock \showarticletitle{Prop: Pre-training with representative words
  prediction for ad-hoc retrieval}. In \bibinfo{booktitle}{\emph{Proceedings of
  the 14th ACM WSDM}}. \bibinfo{pages}{283--291}.
\newblock


\bibitem[\protect\citeauthoryear{Ma, Guo, Zhang, Fan, Li, and Cheng}{Ma
  et~al\mbox{.}}{2021c}]%
        {ma2021b}
\bibfield{author}{\bibinfo{person}{Xinyu Ma}, \bibinfo{person}{Jiafeng Guo},
  \bibinfo{person}{Ruqing Zhang}, \bibinfo{person}{Yixing Fan},
  \bibinfo{person}{Yingyan Li}, {and} \bibinfo{person}{Xueqi Cheng}.}
  \bibinfo{year}{2021}\natexlab{c}.
\newblock \showarticletitle{B-PROP: bootstrapped pre-training with
  representative words prediction for ad-hoc retrieval}. In
  \bibinfo{booktitle}{\emph{Proceedings of the 44th International ACM SIGIR}}.
  \bibinfo{pages}{1513--1522}.
\newblock


\bibitem[\protect\citeauthoryear{Ma, Dou, Xu, Zhang, Jiang, Cao, and Wen}{Ma
  et~al\mbox{.}}{2021a}]%
        {ma2021pre}
\bibfield{author}{\bibinfo{person}{Zhengyi Ma}, \bibinfo{person}{Zhicheng Dou},
  \bibinfo{person}{Wei Xu}, \bibinfo{person}{Xinyu Zhang}, \bibinfo{person}{Hao
  Jiang}, \bibinfo{person}{Zhao Cao}, {and} \bibinfo{person}{Ji-Rong Wen}.}
  \bibinfo{year}{2021}\natexlab{a}.
\newblock \showarticletitle{Pre-training for Ad-hoc Retrieval: Hyperlink is
  Also You Need}. In \bibinfo{booktitle}{\emph{Proceedings of the 30th ACM
  CIKM}}. \bibinfo{pages}{1212--1221}.
\newblock


\bibitem[\protect\citeauthoryear{MacAvaney, Yates, Cohan, and
  Goharian}{MacAvaney et~al\mbox{.}}{2019}]%
        {macavaney2019cedr}
\bibfield{author}{\bibinfo{person}{Sean MacAvaney}, \bibinfo{person}{Andrew
  Yates}, \bibinfo{person}{Arman Cohan}, {and} \bibinfo{person}{Nazli
  Goharian}.} \bibinfo{year}{2019}\natexlab{}.
\newblock \showarticletitle{CEDR: Contextualized embeddings for document
  ranking}. In \bibinfo{booktitle}{\emph{Proceedings of the 42nd International
  ACM SIGIR}}. \bibinfo{pages}{1101--1104}.
\newblock


\bibitem[\protect\citeauthoryear{Mackenzie, Petri, and Moffat}{Mackenzie
  et~al\mbox{.}}{2021}]%
        {mackenzie2021sensitivity}
\bibfield{author}{\bibinfo{person}{Joel Mackenzie}, \bibinfo{person}{Matthias
  Petri}, {and} \bibinfo{person}{Alistair Moffat}.}
  \bibinfo{year}{2021}\natexlab{}.
\newblock \showarticletitle{A Sensitivity Analysis of the MSMARCO Passage
  Collection}.
\newblock \bibinfo{journal}{\emph{arXiv preprint arXiv:2112.03396}}
  (\bibinfo{year}{2021}).
\newblock


\bibitem[\protect\citeauthoryear{Matveeva, Burges, Burkard, Laucius, and
  Wong}{Matveeva et~al\mbox{.}}{2006}]%
        {matveeva2006high}
\bibfield{author}{\bibinfo{person}{Irina Matveeva}, \bibinfo{person}{Chris
  Burges}, \bibinfo{person}{Timo Burkard}, \bibinfo{person}{Andy Laucius},
  {and} \bibinfo{person}{Leon Wong}.} \bibinfo{year}{2006}\natexlab{}.
\newblock \showarticletitle{High accuracy retrieval with multiple nested
  ranker}. In \bibinfo{booktitle}{\emph{Proceedings of the 29th ACM SIGIR}}.
  \bibinfo{pages}{437--444}.
\newblock


\bibitem[\protect\citeauthoryear{Nguyen, Rosenberg, Song, Gao, Tiwary,
  Majumder, and Deng}{Nguyen et~al\mbox{.}}{2016}]%
        {nguyen2016ms}
\bibfield{author}{\bibinfo{person}{Tri Nguyen}, \bibinfo{person}{Mir
  Rosenberg}, \bibinfo{person}{Xia Song}, \bibinfo{person}{Jianfeng Gao},
  \bibinfo{person}{Saurabh Tiwary}, \bibinfo{person}{Rangan Majumder}, {and}
  \bibinfo{person}{Li Deng}.} \bibinfo{year}{2016}\natexlab{}.
\newblock \showarticletitle{MS MARCO: A human generated machine reading
  comprehension dataset}. In \bibinfo{booktitle}{\emph{CoCo@ NIPS}}.
\newblock


\bibitem[\protect\citeauthoryear{Nogueira and Cho}{Nogueira and Cho}{2019}]%
        {nogueira2019passage}
\bibfield{author}{\bibinfo{person}{Rodrigo Nogueira} {and}
  \bibinfo{person}{Kyunghyun Cho}.} \bibinfo{year}{2019}\natexlab{}.
\newblock \showarticletitle{Passage Re-ranking with BERT}.
\newblock \bibinfo{journal}{\emph{arXiv preprint arXiv:1901.04085}}
  (\bibinfo{year}{2019}).
\newblock


\bibitem[\protect\citeauthoryear{Oh~Song, Xiang, Jegelka, and Savarese}{Oh~Song
  et~al\mbox{.}}{2016}]%
        {oh2016deep}
\bibfield{author}{\bibinfo{person}{Hyun Oh~Song}, \bibinfo{person}{Yu Xiang},
  \bibinfo{person}{Stefanie Jegelka}, {and} \bibinfo{person}{Silvio Savarese}.}
  \bibinfo{year}{2016}\natexlab{}.
\newblock \showarticletitle{Deep metric learning via lifted structured feature
  embedding}. In \bibinfo{booktitle}{\emph{Proceedings of the IEEE conference
  on computer vision and pattern recognition}}. \bibinfo{pages}{4004--4012}.
\newblock


\bibitem[\protect\citeauthoryear{Ovaisi, Ahsan, Zhang, Vasilaky, and
  Zheleva}{Ovaisi et~al\mbox{.}}{2020}]%
        {ovaisi2020correcting}
\bibfield{author}{\bibinfo{person}{Zohreh Ovaisi}, \bibinfo{person}{Ragib
  Ahsan}, \bibinfo{person}{Yifan Zhang}, \bibinfo{person}{Kathryn Vasilaky},
  {and} \bibinfo{person}{Elena Zheleva}.} \bibinfo{year}{2020}\natexlab{}.
\newblock \showarticletitle{Correcting for selection bias in learning-to-rank
  systems}. In \bibinfo{booktitle}{\emph{Proceedings of The Web Conference
  2020}}. \bibinfo{pages}{1863--1873}.
\newblock


\bibitem[\protect\citeauthoryear{Prakash, Killingback, and Zamani}{Prakash
  et~al\mbox{.}}{2021}]%
        {prakash2021learning}
\bibfield{author}{\bibinfo{person}{Prafull Prakash}, \bibinfo{person}{Julian
  Killingback}, {and} \bibinfo{person}{Hamed Zamani}.}
  \bibinfo{year}{2021}\natexlab{}.
\newblock \showarticletitle{Learning Robust Dense Retrieval Models from
  Incomplete Relevance Labels}. In \bibinfo{booktitle}{\emph{Proceedings of the
  44th International ACM SIGIR}}. \bibinfo{pages}{1728--1732}.
\newblock


\bibitem[\protect\citeauthoryear{Qin and Liu}{Qin and Liu}{2013}]%
        {qin2013introducing}
\bibfield{author}{\bibinfo{person}{Tao Qin} {and} \bibinfo{person}{Tie-Yan
  Liu}.} \bibinfo{year}{2013}\natexlab{}.
\newblock \showarticletitle{Introducing LETOR 4.0 datasets}.
\newblock \bibinfo{journal}{\emph{arXiv preprint arXiv:1306.2597}}
  (\bibinfo{year}{2013}).
\newblock


\bibitem[\protect\citeauthoryear{Qiu, Li, Qu, Chen, She, Liu, Wu, and Wang}{Qiu
  et~al\mbox{.}}{2022}]%
        {qiu2022dureader_retrieval}
\bibfield{author}{\bibinfo{person}{Yifu Qiu}, \bibinfo{person}{Hongyu Li},
  \bibinfo{person}{Yingqi Qu}, \bibinfo{person}{Ying Chen},
  \bibinfo{person}{Qiaoqiao She}, \bibinfo{person}{Jing Liu},
  \bibinfo{person}{Hua Wu}, {and} \bibinfo{person}{Haifeng Wang}.}
  \bibinfo{year}{2022}\natexlab{}.
\newblock \showarticletitle{DuReader\_retrieval: A Large-scale Chinese
  Benchmark for Passage Retrieval from Web Search Engine}.
\newblock \bibinfo{journal}{\emph{arXiv preprint arXiv:2203.10232}}
  (\bibinfo{year}{2022}).
\newblock


\bibitem[\protect\citeauthoryear{Ren, Qu, Liu, Zhao, She, Wu, Wang, and
  Wen}{Ren et~al\mbox{.}}{2021}]%
        {ren2021rocketqav2}
\bibfield{author}{\bibinfo{person}{Ruiyang Ren}, \bibinfo{person}{Yingqi Qu},
  \bibinfo{person}{Jing Liu}, \bibinfo{person}{Wayne~Xin Zhao},
  \bibinfo{person}{Qiaoqiao She}, \bibinfo{person}{Hua Wu},
  \bibinfo{person}{Haifeng Wang}, {and} \bibinfo{person}{Ji-Rong Wen}.}
  \bibinfo{year}{2021}\natexlab{}.
\newblock \showarticletitle{RocketQAv2: A Joint Training Method for Dense
  Passage Retrieval and Passage Re-ranking}.
\newblock \bibinfo{journal}{\emph{arXiv preprint arXiv:2110.07367}}
  (\bibinfo{year}{2021}).
\newblock


\bibitem[\protect\citeauthoryear{Robertson, Zaragoza, et~al\mbox{.}}{Robertson
  et~al\mbox{.}}{2009}]%
        {robertson2009probabilistic}
\bibfield{author}{\bibinfo{person}{Stephen Robertson}, \bibinfo{person}{Hugo
  Zaragoza}, {et~al\mbox{.}}} \bibinfo{year}{2009}\natexlab{}.
\newblock \showarticletitle{The probabilistic relevance framework: BM25 and
  beyond}.
\newblock \bibinfo{journal}{\emph{Foundations and Trends{\textregistered} in
  Information Retrieval}} \bibinfo{volume}{3}, \bibinfo{number}{4}
  (\bibinfo{year}{2009}), \bibinfo{pages}{333--389}.
\newblock


\bibitem[\protect\citeauthoryear{Robinson, Chuang, Sra, and Jegelka}{Robinson
  et~al\mbox{.}}{2020}]%
        {robinson2020contrastive}
\bibfield{author}{\bibinfo{person}{Joshua Robinson}, \bibinfo{person}{Ching-Yao
  Chuang}, \bibinfo{person}{Suvrit Sra}, {and} \bibinfo{person}{Stefanie
  Jegelka}.} \bibinfo{year}{2020}\natexlab{}.
\newblock \showarticletitle{Contrastive learning with hard negative samples}.
\newblock \bibinfo{journal}{\emph{arXiv preprint arXiv:2010.04592}}
  (\bibinfo{year}{2020}).
\newblock


\bibitem[\protect\citeauthoryear{Schroff, Kalenichenko, and Philbin}{Schroff
  et~al\mbox{.}}{2015}]%
        {schroff2015facenet}
\bibfield{author}{\bibinfo{person}{Florian Schroff}, \bibinfo{person}{Dmitry
  Kalenichenko}, {and} \bibinfo{person}{James Philbin}.}
  \bibinfo{year}{2015}\natexlab{}.
\newblock \showarticletitle{Facenet: A unified embedding for face recognition
  and clustering}. In \bibinfo{booktitle}{\emph{Proceedings of the IEEE
  conference on computer vision and pattern recognition}}.
  \bibinfo{pages}{815--823}.
\newblock


\bibitem[\protect\citeauthoryear{Spark-Jones}{Spark-Jones}{1975}]%
        {spark1975report}
\bibfield{author}{\bibinfo{person}{Karen Spark-Jones}.}
  \bibinfo{year}{1975}\natexlab{}.
\newblock \showarticletitle{Report on the need for and provision of
  an'ideal'information retrieval test collection}.
\newblock \bibinfo{journal}{\emph{Computer Laboratory}} (\bibinfo{year}{1975}).
\newblock


\bibitem[\protect\citeauthoryear{Sun, Wang, Li, Feng, Chen, Zhang, Tian, Zhu,
  Tian, and Wu}{Sun et~al\mbox{.}}{2019}]%
        {sun2019ernie}
\bibfield{author}{\bibinfo{person}{Yu Sun}, \bibinfo{person}{Shuohuan Wang},
  \bibinfo{person}{Yukun Li}, \bibinfo{person}{Shikun Feng},
  \bibinfo{person}{Xuyi Chen}, \bibinfo{person}{Han Zhang},
  \bibinfo{person}{Xin Tian}, \bibinfo{person}{Danxiang Zhu},
  \bibinfo{person}{Hao Tian}, {and} \bibinfo{person}{Hua Wu}.}
  \bibinfo{year}{2019}\natexlab{}.
\newblock \showarticletitle{Ernie: Enhanced representation through knowledge
  integration}.
\newblock \bibinfo{journal}{\emph{arXiv preprint arXiv:1904.09223}}
  (\bibinfo{year}{2019}).
\newblock


\bibitem[\protect\citeauthoryear{Thakur, Reimers, R{\"u}ckl{\'e}, Srivastava,
  and Gurevych}{Thakur et~al\mbox{.}}{2021}]%
        {thakur2021beir}
\bibfield{author}{\bibinfo{person}{Nandan Thakur}, \bibinfo{person}{Nils
  Reimers}, \bibinfo{person}{Andreas R{\"u}ckl{\'e}}, \bibinfo{person}{Abhishek
  Srivastava}, {and} \bibinfo{person}{Iryna Gurevych}.}
  \bibinfo{year}{2021}\natexlab{}.
\newblock \showarticletitle{BEIR: A Heterogenous Benchmark for Zero-shot
  Evaluation of Information Retrieval Models}.
\newblock \bibinfo{journal}{\emph{arXiv preprint arXiv:2104.08663}}
  (\bibinfo{year}{2021}).
\newblock


\bibitem[\protect\citeauthoryear{Vaswani, Shazeer, Parmar, Uszkoreit, Jones,
  Gomez, Kaiser, and Polosukhin}{Vaswani et~al\mbox{.}}{2017}]%
        {vaswani2017attention}
\bibfield{author}{\bibinfo{person}{Ashish Vaswani}, \bibinfo{person}{Noam
  Shazeer}, \bibinfo{person}{Niki Parmar}, \bibinfo{person}{Jakob Uszkoreit},
  \bibinfo{person}{Llion Jones}, \bibinfo{person}{Aidan~N Gomez},
  \bibinfo{person}{{\L}ukasz Kaiser}, {and} \bibinfo{person}{Illia
  Polosukhin}.} \bibinfo{year}{2017}\natexlab{}.
\newblock \showarticletitle{Attention is all you need}. In
  \bibinfo{booktitle}{\emph{Advances in neural information processing
  systems}}. \bibinfo{pages}{5998--6008}.
\newblock


\bibitem[\protect\citeauthoryear{Wang, Bendersky, Metzler, and Najork}{Wang
  et~al\mbox{.}}{2016}]%
        {wang2016learning}
\bibfield{author}{\bibinfo{person}{Xuanhui Wang}, \bibinfo{person}{Michael
  Bendersky}, \bibinfo{person}{Donald Metzler}, {and} \bibinfo{person}{Marc
  Najork}.} \bibinfo{year}{2016}\natexlab{}.
\newblock \showarticletitle{Learning to rank with selection bias in personal
  search}. In \bibinfo{booktitle}{\emph{Proceedings of the 39th International
  ACM SIGIR}}. \bibinfo{pages}{115--124}.
\newblock


\bibitem[\protect\citeauthoryear{Webber and Park}{Webber and Park}{2009}]%
        {webber2009score}
\bibfield{author}{\bibinfo{person}{William Webber} {and}
  \bibinfo{person}{Laurence~AF Park}.} \bibinfo{year}{2009}\natexlab{}.
\newblock \showarticletitle{Score adjustment for correction of pooling bias}.
  In \bibinfo{booktitle}{\emph{Proceedings of the 32nd international ACM
  SIGIR}}. \bibinfo{pages}{444--451}.
\newblock


\bibitem[\protect\citeauthoryear{Xiong, Xiong, Li, Tang, Liu, Bennett, Ahmed,
  and Overwijk}{Xiong et~al\mbox{.}}{2020}]%
        {xiong2020approximate}
\bibfield{author}{\bibinfo{person}{Lee Xiong}, \bibinfo{person}{Chenyan Xiong},
  \bibinfo{person}{Ye Li}, \bibinfo{person}{Kwok-Fung Tang},
  \bibinfo{person}{Jialin Liu}, \bibinfo{person}{Paul Bennett},
  \bibinfo{person}{Junaid Ahmed}, {and} \bibinfo{person}{Arnold Overwijk}.}
  \bibinfo{year}{2020}\natexlab{}.
\newblock \showarticletitle{Approximate nearest neighbor negative contrastive
  learning for dense text retrieval}.
\newblock \bibinfo{journal}{\emph{arXiv preprint arXiv:2007.00808}}
  (\bibinfo{year}{2020}).
\newblock


\bibitem[\protect\citeauthoryear{Yue, Patel, and Roehrig}{Yue
  et~al\mbox{.}}{2010}]%
        {yue2010beyond}
\bibfield{author}{\bibinfo{person}{Yisong Yue}, \bibinfo{person}{Rajan Patel},
  {and} \bibinfo{person}{Hein Roehrig}.} \bibinfo{year}{2010}\natexlab{}.
\newblock \showarticletitle{Beyond position bias: Examining result
  attractiveness as a source of presentation bias in clickthrough data}. In
  \bibinfo{booktitle}{\emph{Proceedings of the 19th international conference on
  World wide web}}. \bibinfo{pages}{1011--1018}.
\newblock


\bibitem[\protect\citeauthoryear{Zhan, Mao, Liu, Guo, Zhang, and Ma}{Zhan
  et~al\mbox{.}}{2021}]%
        {zhan2021optimizing}
\bibfield{author}{\bibinfo{person}{Jingtao Zhan}, \bibinfo{person}{Jiaxin Mao},
  \bibinfo{person}{Yiqun Liu}, \bibinfo{person}{Jiafeng Guo},
  \bibinfo{person}{Min Zhang}, {and} \bibinfo{person}{Shaoping Ma}.}
  \bibinfo{year}{2021}\natexlab{}.
\newblock \showarticletitle{Optimizing Dense Retrieval Model Training with Hard
  Negatives}.
\newblock \bibinfo{journal}{\emph{arXiv preprint arXiv:2104.08051}}
  (\bibinfo{year}{2021}).
\newblock


\bibitem[\protect\citeauthoryear{Zhan, Mao, Liu, Zhang, and Ma}{Zhan
  et~al\mbox{.}}{2020}]%
        {zhan2020repbert}
\bibfield{author}{\bibinfo{person}{Jingtao Zhan}, \bibinfo{person}{Jiaxin Mao},
  \bibinfo{person}{Yiqun Liu}, \bibinfo{person}{Min Zhang}, {and}
  \bibinfo{person}{Shaoping Ma}.} \bibinfo{year}{2020}\natexlab{}.
\newblock \showarticletitle{RepBERT: Contextualized text embeddings for
  first-stage retrieval}.
\newblock \bibinfo{journal}{\emph{arXiv preprint arXiv:2006.15498}}
  (\bibinfo{year}{2020}).
\newblock


\bibitem[\protect\citeauthoryear{Zhang, Gong, Shen, Lv, Duan, and Chen}{Zhang
  et~al\mbox{.}}{2021}]%
        {zhang2021adversarial}
\bibfield{author}{\bibinfo{person}{Hang Zhang}, \bibinfo{person}{Yeyun Gong},
  \bibinfo{person}{Yelong Shen}, \bibinfo{person}{Jiancheng Lv},
  \bibinfo{person}{Nan Duan}, {and} \bibinfo{person}{Weizhu Chen}.}
  \bibinfo{year}{2021}\natexlab{}.
\newblock \showarticletitle{Adversarial Retriever-Ranker for dense text
  retrieval}.
\newblock \bibinfo{journal}{\emph{arXiv preprint arXiv:2110.03611}}
  (\bibinfo{year}{2021}).
\newblock


\bibitem[\protect\citeauthoryear{Zhang and Yang}{Zhang and Yang}{2021}]%
        {zhang2021dml}
\bibfield{author}{\bibinfo{person}{Xuanyu Zhang} {and} \bibinfo{person}{Qing
  Yang}.} \bibinfo{year}{2021}\natexlab{}.
\newblock \showarticletitle{DML: Dynamic Multi-Granularity Learning for
  BERT-Based Document Reranking}. In \bibinfo{booktitle}{\emph{Proceedings of
  the 30th ACM CIKM}}. \bibinfo{pages}{3642--3646}.
\newblock


\end{thebibliography}

\end{document}